\PassOptionsToPackage{unicode}{hyperref}
\PassOptionsToPackage{hyphens}{url}
\PassOptionsToPackage{dvipsnames,svgnames,x11names}{xcolor}
\documentclass[
  sn-basic,
  10pt,
]{sn-jnl}

\usepackage{xcolor}
\usepackage{amsmath,amssymb}
\usepackage{pgfplots}
\pgfplotsset{compat=1.18}
\usepackage{tikz}
\usepackage{iftex}
\ifPDFTeX
  \usepackage[T1]{fontenc}
  \usepackage[utf8]{inputenc}
  \usepackage{textcomp} 
\else 
  \usepackage{unicode-math} 
  \defaultfontfeatures{Scale=MatchLowercase}
  \defaultfontfeatures[\rmfamily]{Ligatures=TeX,Scale=1}
\fi
\usepackage{lmodern}
\ifPDFTeX\else
\fi
\IfFileExists{upquote.sty}{\usepackage{upquote}}{}
\IfFileExists{microtype.sty}{
  \usepackage[]{microtype}
  \UseMicrotypeSet[protrusion]{basicmath} 
}{}
\makeatletter
\ifx\paragraph\undefined\else
  \let\oldparagraph\paragraph
  \renewcommand{\paragraph}{
    \@ifstar
      \xxxParagraphStar
      \xxxParagraphNoStar
  }
  \newcommand{\xxxParagraphStar}[1]{\oldparagraph*{#1}\mbox{}}
  \newcommand{\xxxParagraphNoStar}[1]{\oldparagraph{#1}\mbox{}}
\fi
\ifx\subparagraph\undefined\else
  \let\oldsubparagraph\subparagraph
  \renewcommand{\subparagraph}{
    \@ifstar
      \xxxSubParagraphStar
      \xxxSubParagraphNoStar
  }
  \newcommand{\xxxSubParagraphStar}[1]{\oldsubparagraph*{#1}\mbox{}}
  \newcommand{\xxxSubParagraphNoStar}[1]{\oldsubparagraph{#1}\mbox{}}
\fi
\makeatother

\usepackage{longtable,booktabs,array}
\usepackage{calc} 
\usepackage{etoolbox}
\makeatletter
\patchcmd\longtable{\par}{\if@noskipsec\mbox{}\fi\par}{}{}
\makeatother
\IfFileExists{footnotehyper.sty}{\usepackage{footnotehyper}}{\usepackage{footnote}}
\makesavenoteenv{longtable}
\usepackage{graphicx}
\makeatletter
\newsavebox\pandoc@box
\newcommand*\pandocbounded[1]{
  \sbox\pandoc@box{#1}%
  \Gscale@div\@tempa{\textheight}{\dimexpr\ht\pandoc@box+\dp\pandoc@box\relax}%
  \Gscale@div\@tempb{\linewidth}{\wd\pandoc@box}%
  \ifdim\@tempb\p@<\@tempa\p@\let\@tempa\@tempb\fi
  \ifdim\@tempa\p@<\p@\scalebox{\@tempa}{\usebox\pandoc@box}%
  \else\usebox{\pandoc@box}%
  \fi%
}
\def\fps@figure{htbp}
\makeatother

\providecommand{\tightlist}{%
  \setlength{\itemsep}{0pt}\setlength{\parskip}{0pt}}

\usepackage{graphicx}%
\usepackage{multirow}%
\usepackage{amsmath,amssymb,amsfonts}%
\usepackage{amsthm}
\theoremstyle{thmstyleone}
\newtheorem{proposition}{Proposition}[section]
\theoremstyle{thmstyleone}
\newtheorem{theorem}[proposition]{Theorem}
\newtheorem{corollary}[proposition]{Corollary}
\theoremstyle{remark}
\AtBeginDocument{\renewcommand*{\proofname}{Proof}}

\usepackage{mathrsfs}%
\usepackage[title]{appendix}%
\usepackage{xcolor}%
\usepackage{textcomp}%
\usepackage{manyfoot}%
\usepackage{booktabs}%
\usepackage{algorithm}%
\usepackage{algorithmicx}%
\usepackage{algpseudocode}%
\usepackage{listings}%
\usepackage{setspace}%
\usepackage{mathtools}%
\usepackage{enumitem}

\makeatletter
\@ifpackageloaded{caption}{}{\usepackage{caption}}
\AtBeginDocument{%
\ifdefined\contentsname
  \renewcommand*\contentsname{Table of contents}
\else
  \newcommand\contentsname{Table of contents}
\fi
\ifdefined\listfigurename
  \renewcommand*\listfigurename{List of Figures}
\else
  \newcommand\listfigurename{List of Figures}
\fi
\ifdefined\listtablename
  \renewcommand*\listtablename{List of Tables}
\else
  \newcommand\listtablename{List of Tables}
\fi
\ifdefined\figurename
  \renewcommand*\figurename{\textbf{Fig.}}
\else
  \newcommand\figurename{\textbf{Fig.}}
\fi
\ifdefined\tablename
  \renewcommand*\tablename{\sffamily\textbf{Table}}
\else
  \newcommand\tablename{\sffamily\textbf{Table}}
\fi
}
\@ifpackageloaded{float}{}{\usepackage{float}}
\floatstyle{ruled}
\@ifundefined{c@chapter}{\newfloat{codelisting}{h}{lop}}{\newfloat{codelisting}{h}{lop}[chapter]}
\floatname{codelisting}{Listing}

\usepackage{amsthm}
\theoremstyle{plain}
\theoremstyle{plain}
\theoremstyle{remark}
\AtBeginDocument{\renewcommand*{\proofname}{Proof}}
\makeatother
\makeatletter
\@ifpackageloaded{caption}{}{\usepackage{caption}}
\@ifpackageloaded{subcaption}{}{\usepackage{subcaption}}
\makeatother
\usepackage{bookmark}
\IfFileExists{xurl.sty}{\usepackage{xurl}}{} 
\hypersetup{
  pdftitle={A spliced preferential attachment model for degree distributions in networks},
  pdfauthor={Thomas William Boughen; Clement Lee; Vianey Palacios Ramirez},
  pdfkeywords={Network degrees - Discrete extremes - Power law -
Preferential attachment},
  colorlinks=true,
  linkcolor={blue},
  filecolor={Maroon},
  citecolor={Blue},
  urlcolor={Blue},
  pdfcreator={LaTeX via pandoc}}

\title[A spliced preferential attachment model for degree distributions in networks]{A spliced preferential attachment model for degree distributions in networks}

\author*[1]{\fnm{Thomas William} \sur{Boughen}}\email{t.w.boughen1@newcastle.ac.uk}\author[1]{\fnm{Clement} \sur{Lee}}\author[1]{\fnm{Vianey Palacios} \sur{Ramirez}}
\affil[1]{\orgdiv{School of Mathematics, Statistics and
Physics}, \orgname{Newcastle University}}

\abstract{Identifying the generating mechanism of a network is
challenging as, more often than not, only snapshots are available, but
not the full evolution. One candidate for the generating mechanism is the general
preferential attachment (GPA), in which existing nodes gain new
connections at a rate governed by a preference function of their current
degree, which, in its simplest form, results in a degree
distribution that follows the power law. 
However, the ubiquity of the power law in real-life networks has been challenged on two fronts: alternative distributions often fit comparably well, and recent works using extreme value methods have shown that the tail of the degree distribution, while still regularly varying, tends to be lighter than the body implies.
In this paper, we propose a GPA model with a flexible preference function. Using methods for discrete extremes, we characterise the tail behaviour of the limiting
degree distribution directly by the preference function.
This direct connection facilitates the inference of the model parameters using snapshot data alone, and sidesteps the need of traditional threshold-based extreme value methods, which lack interpretability and suffer from identifiability issues. Comprehensive simulation studies show that our model recovers the parameters well, while
applications to real-life networks demonstrate comparable performance to established alternatives and provide insights into the growth dynamics of the networks.}


\pacs[\sffamily AMS 2000 Subject Classification]{05C80· 60J85· 62G32}

\begin{document}
\maketitle

\newpage

\section{Introduction}\label{sec-introduction}

Networks have become powerful tools for representing and analysing
complex systems across a wide range of fields. In network
science and statistics, they have been studied by various families of
models, from stochastic block models for detecting communities online
\citep{Latouche11}, to exponential random graph models (ERGMs) for
analysing the global trade network \citep{Setayesh22}, and mechanistic
models for investigating patterns in neural systems \citep{Betzel17}.

The \textit{degree} of a node in a network, that is, the number of connections it has to other nodes, is one of its most basic and widely studied properties. A network is said to be \textit{scale-free} if the proportion of nodes with degree greater than $k$ is proportional to $k^{-\gamma}$ for large $k$, that is, if its degree distribution has a regularly varying tail with tail index $\gamma$. The study of this tail behaviour sits at the intersection of network science and extreme value theory (EVT), and whether scale-freeness is in fact ubiquitous in real networks has been the subject of considerable debate, in part because degree distributions never follow the power law exactly, different definitions of scale-freeness are used across the literature, and the power law is tested against different competing alternatives. 

Critiquing ubiquity, \citet{Broido_2019} evaluated power-law fits across nearly a thousand networks, finding strong evidence for scale-freeness in only four percent of cases. Conversely, defining scale-freeness via regular variation, \citet{Voitalov_2019} argue that scale-free networks are far more prevalent once deviations from a pure power law are accounted for. Building on this, \citet{Lee24} show that many networks are partially scale-free: while the body of the degree distribution follows a power law, the extreme tail is lighter than what the body implies, though still regularly varying.

This body of work motivates treating the tail of the degree distribution as a distinct object of inference rather than assuming a single power law throughout, an idea closely related to bulk-tail models in the extremes literature, such as the extended generalised Pareto distribution (EGPD) of \citet{Papastathopoulos2013} and \citet{Naveau2016} (see also \cite{HandbookExtremesCh05}), which jointly model low, moderate, and extreme values without a hard threshold. As we clarify below, the model we propose in this paper, which we call the \emph{spliced GPA model}, offers a mechanistic counterpart to this idea.

Understanding a network requires learning not just about the static degree distribution, but about the dynamic growth mechanisms that generate it, an area where our model pioneers a direct connection between growth rules and extreme value theory. While \cite{Voitalov_2019} and \cite{Lee24} have incorporated methods for extreme values in the network analyses, they remain largely descriptive as no information about the growth of the networks is revealed. On the other hand, growth rules are explicitly stipulated in the model by \cite{Barabasi99} that popularised using the power law for degrees, and its generalisation, the general preferential attachment (GPA) model. As new nodes join the network, an existing node with degree $k$ gains edges at a rate proportional to a non-negative preference function $b(k)$. \cite{Barabasi99} showed that, when $b(k) = k + 1$, in the limit the resulting degree distribution is regularly varying with tail index 2.  Subsequently, if a real network is deemed scale-free, one could postulate that preferential attachment is the underlying growth mechanism.

The Barab\'asi-Albert (BA) model provided the foundations for various
generalisations. \citet{krapivsky01} considered
\(b(k) = (k+1)^\alpha\), and showed that the tail of the degree
distribution is not regularly varying (and therefore not following the
power law) when \(0<\alpha<1\), and when \(\alpha>1\) a finite number of
nodes end up with all edges after a certain point, resulting in a
degenerate degree distribution. \citet{rudas07} followed in the footsteps of \citet{krapivsky01}, by
considering a GPA tree and using theory from continuous branching
processes, deriving a limiting degree distribution in terms of the
preference function \(b(k)\). \citet{wang2022random} returned to a
linear preference function of the form \(b(k) = k+\varepsilon~(\varepsilon>0)\) and incorporated edge reciprocity, resulting in the joint
distribution of in-degree and out-degree being multivariate regularly
varying and having the property of hidden regular variation.
Unfortunately, research in this area
tends to only focus on the theoretical asymptotic results of network
growth models, with little analysis of real networks.

This paper aims to address simultaneously the gap between the applied and theoretical
works, and the issue with the descriptive nature of existing EVT-based methods. Specifically, we ask, if a network is assumed to come from a GPA model, can
we use the degree distribution to directly infer the parameters of
the preference function and learn about the growth mechanisms?
Moreover, proper consideration is given to the tail of the degree
distribution, because otherwise the effects of the largest degrees,
which correspond to the most influential nodes, deviating from the power law will be discounted.
To this end, we propose a spliced GPA model with a flexible, piecewise preference function that behaves sub- or super-linearly for low degrees, and linearly beyond a threshold. The bulk-tail behaviour of the resulting degree distribution is a \emph{derived consequence} of this mechanistic assumption about how the network grows, so that the fitted parameters retain a direct interpretation in terms of network growth dynamics.

Establishing this direct connection between dynamic growth mechanics and limiting tail behaviour relies fundamentally on EVT. However, applying EVT in this context presents an immediate challenge due to the discrete nature of network degrees, as many standard discrete distributions do not satisfy the conditions required to belong to a maximum domain of attraction (MDA). \citet{shimura12}
showed that a discrete distribution \(F\) must be long-tailed, that is
\(\lim_{x\rightarrow\infty}\bar F(x+1)/\bar F(x) = 1\), to be
in an MDA, and introduced the notion of recoverability to the MDA, also
referred to as the discrete MDA \citep{hitz24}. Using the
theoretical guarantees of being in a discrete MDA for a discrete
distribution by \citet{shimura12}, we demonstrate how the tail of the
degree distribution is directly influenced by \(b(k)\), confirming that asymptotic linearity of $b(k)$ is precisely the condition under which the limiting degree distribution is regularly varying. Subsequently, we propose a class of preference functions that not only imply
regularly varying degree distributions but are also realistic for
real networks in terms of tail behaviour. These analytical results enable the likelihood of the
degree distribution to be expressed in terms of the parameters of
\(b(k)\), thus allowing the underlying growth mechanism of the
network to be inferred directly.

The remainder of the paper is as follows: Section~\ref{sec-tail} introduces the GPA model and the theoretical results
for its limiting degree distribution, focusing on the characterisation of the tail behaviour by the preference function
\(b(k)\), before studying our proposed model, which is a GPA model with a specific preference function.
The simulation study in
Section~\ref{sec-sim} demonstrates that the parameters can be recovered
from fitting the model to the degrees alone and as a
consequence recover the underlying growth mechanism.
Section~\ref{sec-real} fits the model to real data and provides
posterior estimates for the preference function. Section~\ref{sec-conc}
provides a discussion of this paper and avenues for future
work.

\section{Spliced General Preferential Attachment Model}\label{sec-tail}

In the first half of this section, we study the general preferential attachment (GPA) model, by deriving its limiting degree distribution and characterising its tail behaviour. In the second half, we look at the proposed model, which is a particular case of the GPA model, by examining its numerical properties and outlining the inference procedure.

\subsection{From a branching process to the limiting degree distribution}

Following \citet{rudas07}, the GPA model is defined as follows:
starting at time \(t=0\) with an initial network of \(m\) nodes that
each have no edges, at times \(t=1,2,\ldots\) a new node is added to
the network bringing with it \(m\) directed edges from the new node;
the targets for each of these edges are selected from the nodes
already in the network with weights proportional to some non-decreasing
preference function \(b(\cdot)\) of their degree, where
\(b: \mathbb N \mapsto \mathbb R^+\setminus\{0\}\) is such that:

\begin{equation}\phantomsection\label{eq-condb2}{
\sum_{k=0}^\infty\frac{1}{b(k)} = \infty.
}\end{equation}
Special cases of this model include:
\begin{enumerate}[label=(\alph*)]
\item the BA model when $b(k) = k+1$, which in the limit of $t\rightarrow \infty$ leads to a power-law degree distribution with tail index 2, and 
\item the uniform attachment model when $b(k)=c\:(>0)$, which in the limit of $t\rightarrow \infty$ leads to a degree distribution with a tail that is not regularly varying.
\end{enumerate}
Condition~\eqref{eq-condb2} and monotonicity are assumed for $b(k)$ throughout this paper.

As the survival function of the limiting degree distribution, called the limiting survival hereafter, is crucial to our proposed model in Section~\ref{sect.splice-gpa}, its analytical derivations in the case where $m=1$ are outlined below.

Consider a continuous time branching process \(\zeta(t)\) driven by a
Markovian pure birth process, with \(\zeta(0)=0\) and birth rates
depending on a non-negative function \(b(\cdot)\): \[
\Pr(\zeta(t+\text{d}t)=k+1|\zeta(t)=k) = b(k)\text{d}t + o(\text{d}t).
\] 
Let $\rho:[0,\infty)\rightarrow (0,\infty)$ be the density of the point process associated with $\zeta(t)$, describing the growth of an individual node, and let
\begin{equation*}\label{eq-laplace}\hat \rho(\lambda)\coloneqq \int_0^\infty e^{-\lambda t}\rho(t)\text{d}t.\end{equation*}
denote its Laplace transform. The Laplace transform can be easily computed, given the fact that the intervals between successive jumps of $\zeta(t)$ are independent exponentially distributed random variables with parameters $b(0),b(1),\dots,$ respectively: 
\begin{equation}\label{eq-laplace-condition}
\hat \rho(\lambda)=\sum_{n=0}^\infty\prod_{i=0}^{n-1}\frac{b(i)}{\lambda+b(i)}
\end{equation}
Condition~\eqref{eq-condb2} implies that the equation 
$\hat\rho(\lambda) = 1$ has a unique root $\lambda^*$, which is the so-called Malthusian parameter.

We now construct the tree $\Upsilon(t)$ generated by $\zeta(t)$:
$\Upsilon(0)=\{\emptyset\}$, and each node $x\in\Upsilon(t)$
independently produces a new child at rate $b(\deg(x,\Upsilon(t)))$,
where $\deg(x,\Upsilon(t))$ is the degree of $x$ in $\Upsilon(t)$ at
time $t$. We write $\Upsilon(t)_{\downarrow x}$ for the subtree of
$\Upsilon(t)$ rooted at $x$, and $|\Upsilon(t)|$ for the total number
of nodes at time $t$.
 
\citet{rudas07} show that, for $m=1$, the GPA model is equivalent in distribution to the branching process constructed above.
This equivalence lets us express the limiting survival as the empirical proportion of nodes in
$\Upsilon(t)$ whose degree (once re-rooted at themselves) exceeds a
threshold $k\in\mathbb N$:
\begin{equation}\label{eq-empirical-limit}
\bar F(k) \coloneqq \lim_{t\to\infty}
\frac{\sum_{x\in\Upsilon(t)}\mathbb I\{\deg(x,\Upsilon(t)_{\downarrow x})>k\}}
{|\Upsilon(t)|}.
\end{equation}

As every node's degree process relative to its own subtree is an independent copy of $\zeta(t)$  Theorem 1 of \citet{rudas07} shows that the limit in Equation~\eqref{eq-empirical-limit} reduces to a single expectation evaluated at the root $x$. Full details on this construction, along with further derivations providing visual intuition, are provided in Section~1 of the Supplementary Material. Specifically, this reduction yields
\begin{equation}\label{eq-rudas}
\bar F(k) = \lambda^*\int_0^\infty e^{-\lambda^* t}\,
\mathbb E\big[\mathbb I\{\deg(x,\Upsilon(t))>k\}\big]\,\mathrm dt.
\end{equation}
By construction, the degree of the root grows according to the pure birth process generating $\Upsilon(t)$, so $\deg(x,\Upsilon(t))$ coincides path-wise with $\zeta(t)$. This allows us to compute the closed-form survival function:
\begin{equation}\label{eq-surv-origin}
\bar F(k) = \lambda^*\int_0^\infty e^{-\lambda^* t}\,
\mathbb E\big[\mathbb I\{\deg(x,\Upsilon(t))>k\}\big]\,\mathrm dt
= \prod_{i=0}^k\frac{b(i)}{\lambda^*+b(i)}.
\end{equation}

Therefore, the corresponding probability mass function (pmf) of the limiting degree
distribution \(f(k) = \bar F(k-1) - \bar F(k)\) is

\begin{equation}\phantomsection\label{eq-pmf-origin}{
f(k) = \frac{\lambda^*}{\lambda^* + b(k)}\prod_{i=0}^{k-1}\frac{b(i)}{\lambda^*+b(i)}.
}\end{equation}
While these foundational derivations originate from \citet{rudas07}, the closed-form limiting survival function in Equation~\eqref{eq-surv-origin} serves as our primary tool and starting point for developing the main theoretical results established in the remainder of this paper.
\subsection{Tail behaviour of limiting degree distribution}\label{sect.splice-gpa}

In this subsection, we motivate the \textit{spliced} GPA model with a specific form of $b(k)$, and provide the main analytical results of this paper, which are then applied to the proposed model to examine the tail behaviour of its limiting degree distribution.

\begin{figure}

\centering{

\includegraphics[width=70mm,height=\textheight,keepaspectratio]{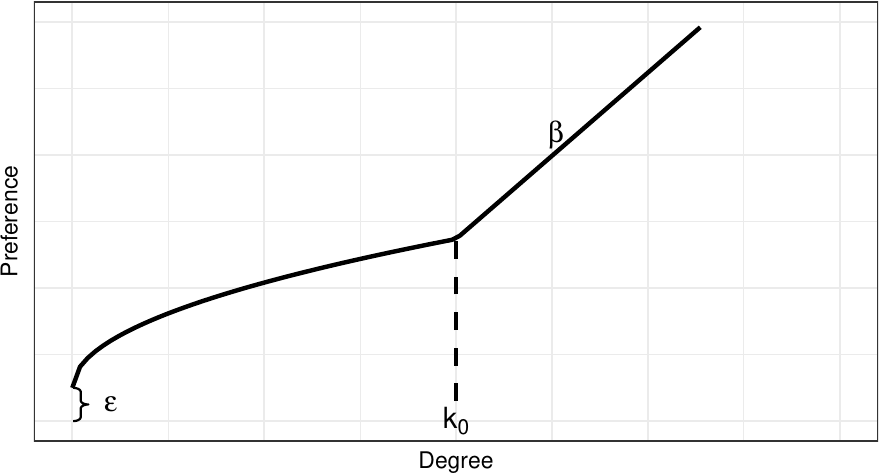}

}

\caption{\label{fig-ex}Schematic of the preference function $b(k)$ of the spliced GPA model}

\end{figure}

Equation~\eqref{eq-surv-origin} implies that the preference function $b(k)$ directly influences the limiting survival, and in turn the tail behaviour of the limiting degree distribution. While $b(k)=(k+1)^{\alpha}$ generalises the BA model naturally, the case where $\alpha\leq1$ implies a rather restrictive tail behaviour, as will be shown by Corollary~\ref{thm-omega2}, while the case where $\alpha>1$ does not satisfy Condition~\eqref{eq-condb2}, and has been shown to lead to degenerative degree distributions \citep{krapivsky01}. Combining these cases and being informed by \citet{Lee24}, we propose a piecewise function:
\begin{equation}\phantomsection\label{eq-pref}{
b(k) = \begin{cases}
k^\alpha + \varepsilon,&k\le k_0,\\
k_0^\alpha + \varepsilon + \beta(k-k_0), &k> k_0
\end{cases}
}\end{equation} for \(\alpha,\beta, \varepsilon>0\) and
\(k_0\in\mathbb N\). 
The schematic of such a spliced preference function is shown in Figure~\ref{fig-ex}. For textual interpretation, we refer to $\alpha$, $\beta$, $\epsilon$, and $k_0$ as the power, slope, zero appeal, and splice point, respectively.

We first characterise the tail behaviour of the limiting degree distribution by the general form of \(b(k)\), before applying the characterisation to our model in Equation~\eqref{eq-pref}.
For a discrete distribution \(F\) with survival function \(\bar F\) and
\(k\in\mathbb Z^+\), a central tool of the characterisation \citep{shimura12} is the quantity
\begin{equation}
\Omega(F,k) := \left(\log\displaystyle\frac{\bar F (k+1)}{\bar F (k+2)}\right)^{-1} - \left(\log\displaystyle\frac{\bar F (k)}{\bar F (k+1)}\right)^{-1}.\label{eq-omega}
\end{equation}
Closely related is the behaviour of $\bar{F}(k+1)/\bar{F}(k)$, for which the limit as $k\rightarrow\infty$ is assumed to exist throughout. Specifically, $F$ is called \textit{long-tailed} if $\lim_{k\rightarrow\infty}\bar{F}(k+1)/\bar{F}(k)=1$, and \textit{not long-tailed} otherwise. \citet{shimura12} established the following results:
\begin{enumerate}
\item If $\lim_{k\rightarrow\infty} \Omega(F,k) = 0$, where $F$ is called \textit{light-tailed}, then its tail behaviour is divided into two cases:
  \begin{itemize}
    \item If $F$ is not long-tailed, then it is \textit{recovered to} the Gumbel MDA.
    \item If $F$ is long-tailed, then it is \textit{in} the Gumbel MDA.
  \end{itemize}
\item If $\lim_{k\rightarrow\infty} \Omega(F,k) = 1/\alpha$ ($\alpha>0$), then $F$ is regularly varying with $\bar F(k) \sim k^{-\alpha}$, and hence belongs to the Fr\'echet MDA with tail index $\alpha$.
\end{enumerate}

The results above offer a natural approach to analysing the tail
behaviour of the limiting degree distribution of the GPA model. We state the main results:

\begin{theorem}\label{thm-coro-omega2}
Let $F(k)$ be the limiting degree distribution of a GPA model with preference function $b(k)$ where $\lim_{k\rightarrow\infty}b(k)<\infty$. Then \[\lim_{k\rightarrow\infty}\Omega(F, k) = 0.\]
Furthermore, \(F\) is recovered to but is not in the Gumbel MDA. See Appendix
\ref{sec-remproof} for the proof.

\label{thm-rem-omega}

\end{theorem}

\begin{theorem}[]\protect\hypertarget{prp-omega}{}\label{thm-prp-omega}

Let \(F (k)\) be the limiting degree distribution of a GPA model with preference function \(b(k)\) where $b(k)\rightarrow\infty$ as $k\rightarrow\infty$. 
Then \[
\lim_{k\rightarrow\infty}\Omega(F,k) = \lim_{k\rightarrow\infty}\frac{b(k+1)-b(k)}{\lambda^*},
\]
where \(\lambda^*\) is the Malthusian parameter of the corresponding
branching process. Furthermore, $F$ is long-tailed. See Appendix~\ref{sec-prpproof} for the proof.

\end{theorem}

\begin{corollary}[]\protect\hypertarget{thm-omega2}{}\label{thm-omega2}

Let $F(k)$ be the limiting degree distribution of a GPA model with preference function \(b(k)\) such that
\(b(k)\rightarrow\infty\) as \(k\rightarrow\infty\). Then:

\begin{itemize}
\item
  $F$ is light-tailed if and only if
  \(\lim_{k\rightarrow\infty}[b(k+1)-b(k)]=0\). Furthermore, \(F\) is in the
  Gumbel MDA as it is also long-tailed.
\item
  $F$ is regularly varying if and only if
  \(\lim_{k\rightarrow\infty}[b(k+1)-b(k)]=c>0\), in which case the tail
  index is \(\lambda^*/c\), where \(\lambda^*\) is the Malthusian
  parameter of the corresponding branching process.
\end{itemize}

\end{corollary}

\begin{proof}
    For the case $\lim_{k\rightarrow\infty}[b(k+1)-b(k)]=0$, Theorem~\ref{thm-prp-omega} implies both $\lim_{k\rightarrow}\Omega(F,k)=0$ and that $F$ is long-tailed.
    
    For the case $\lim_{k\rightarrow\infty}[b(k+1)-b(k)]=c>0$, Theorem~\ref{thm-prp-omega} implies $\lim_{k\rightarrow\infty}\Omega(F,k)=c/\lambda^{*}>0$. According to Theorem~6(iii) of \cite{shimura12}, $F$ is in the Fr\'echet MDA and regularly varying with tail index the reciprocal of $\lim_{k\rightarrow\infty}\Omega(F,k)$, i.e. $\lambda^{*}/c$.
\end{proof}
Applications of Corollary~\ref{thm-omega2} to the following special cases are consistent with previous findings in the literature:
\begin{enumerate}[label=(\alph*)]
\item Uniform attachment: when \(b(k)=c>0\), it is clear that
\(\lim_{k\rightarrow\infty}b(k)<\infty\). By Theorem~\ref{thm-rem-omega}, \(\lim_{k\rightarrow\infty}\Omega(F,k)=0\), and the limiting degree distribution is recovered to (but not in) the Gumbel MDA.
\item Sub-linear preferential attachment \citep{krapivsky01}: when $b(k)=(k+1)^\alpha$ where $\alpha<1$, we have \(\lim_{k\rightarrow\infty}[b(k+1)-b(k)]=0\), implying a light-tailed limiting degree distribution.
\item The BA model \citep{Barabasi99}: when $b(k)=k+1$, we have \(\lim_{k\rightarrow\infty}[b(k+1)-b(k)]=1\), implying a tail index of $\lambda^{*}$, which is equal to $2$ upon solving the associated Malthusian equation.
\end{enumerate}

Theorem~\ref{thm-coro-omega2} and Corollary~\ref{thm-omega2} substantiate the direct connection between the preference function and the tail behaviour of the limiting degree distribution.
Applying them to the spliced GPA model in Equation~\eqref{eq-pref} yields $\lim_{k\rightarrow\infty}[b(k+~1)-b(k)]=\beta$, and the tail index of the limiting degree distribution is $\lambda^{*}/\beta$, where the associated Malthusian parameter $\lambda^{*}$ satisfies Equation~\eqref{eq-laplace-condition} and therefore depends on the model parameters. This implies that the tail index can vary flexibly according to the parameter combination.

\subsection{Numerical properties of spliced GPA model} \label{sec-numerical-properties}

So far, the preference function in Equation~\eqref{eq-pref} simultaneously
allows for the inclusion of sub/super-linear
behaviour below the splice point $k_0$, and ensures regular
variation of the limiting degree distribution, provided that $\lim_{k\rightarrow\infty}[b(k+~1)-b(k)]$ exists and is positive. In this subsection, we examine the spliced GPA model numerically, demonstrate its
flexibility in terms of the tail index of the limiting degree distribution, and provide a connection to existing threshold-based extreme value methods.

To begin with, substituting Equation~\eqref{eq-pref} into Equation~\eqref{eq-surv-origin} yields the limiting survival of the proposed model
\begin{equation}\phantomsection\label{eq-polysurv}{
\bar F(k) = \begin{cases}
\prod_{i=0}^{k}\frac{i^\alpha + \varepsilon}{\lambda^*+i^\alpha + \varepsilon},&k\le k_0,\\
\left(\prod_{i=0}^{k_0-1}\frac{i^\alpha + \varepsilon}{\lambda^*+i^\alpha + \varepsilon}\right)
\frac{\Gamma\left(\frac{\lambda^*+k_0^\alpha + \varepsilon}{\beta}\right)}{\Gamma\left(\frac{k_0^\alpha + \varepsilon}{\beta}\right)} \frac{\Gamma\left(k-k_0 + 1 +\frac{k_0^\alpha + \varepsilon}{\beta}\right)}{\Gamma\left(k-k_0 + 1 +\frac{\lambda^* +k_0^\alpha + \varepsilon}{\beta}\right)}
,&k > k_0,
\end{cases}
}\end{equation} 
with the derivations in Appendix~\ref{sec-gamma}.
The Malthusian parameter \(\lambda^*\) satisfies
\(\hat \rho(\lambda^*)=1\) where Equation~\eqref{eq-laplace-condition} is reduced to:
\begin{equation}\phantomsection\label{eq-rho}{
\hat\rho(\lambda) = \sum_{n=0}^{k_0}\prod_{i=0}^{n-1}\frac{i^\alpha + \varepsilon}{\lambda+i^\alpha + \varepsilon} + \left(\frac{k_0^\alpha + \varepsilon}{\lambda-\beta}\right)\prod_{i=0}^{k_0-1}\frac{i^\alpha + \varepsilon}{\lambda + i^\alpha + \varepsilon}, 
}\end{equation} which has to be solved numerically for most parameter
choices. Also, note that \(\lambda>\beta\). See Appendix
\ref{sec-rhoproof} for the derivations of Equation~\eqref{eq-rho}.

\begin{figure}

\centering{

\includegraphics[width=119mm,height=\textheight,keepaspectratio]{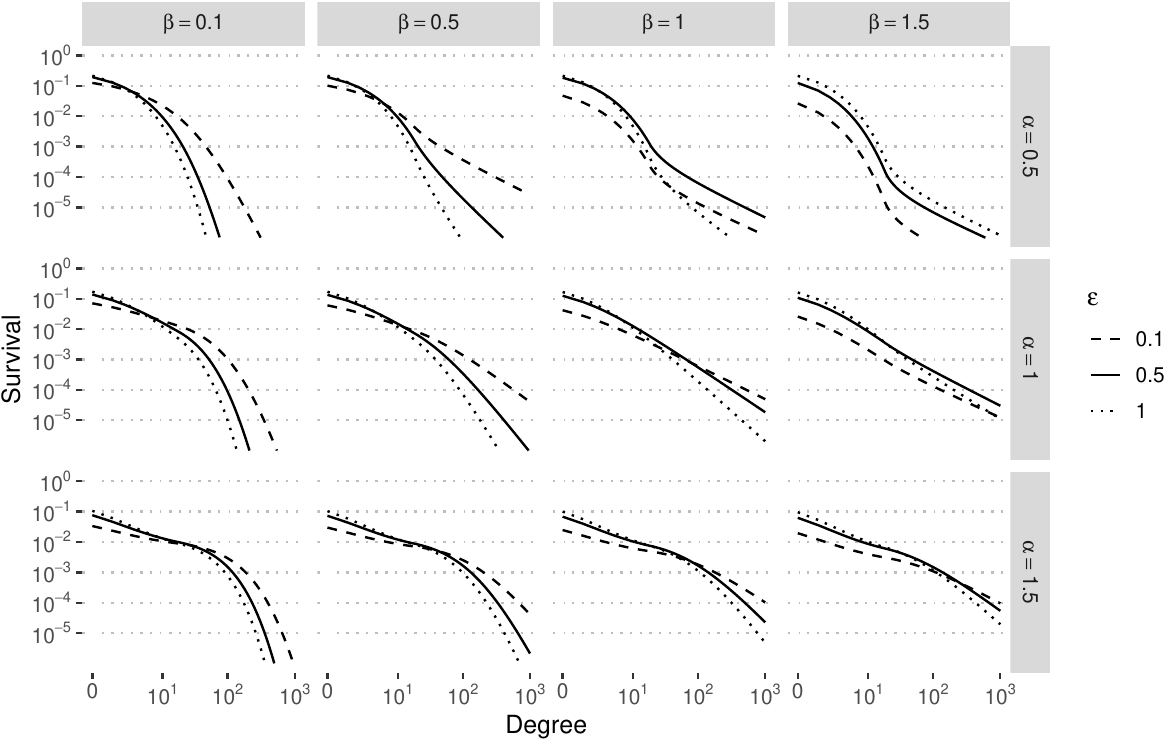}

}

\caption{\label{fig-polylinsurv}The limiting survival, according to
various combinations of \((\alpha, \beta, \varepsilon)\) and \(k_0=20\)
of the spliced GPA model, on the log-log scale}

\end{figure}

As illustrated in Figure~\ref{fig-polylinsurv}, the limiting survival distribution $\bar{F}(k)$ exhibits non-linearity on a log-log scale. The model can capture a wide range of tail behaviours, with tail indices spanning from $1.014$ with $(\alpha,\beta,\varepsilon,k_0)=(1,1.5,0.1,20)$ (middle row, rightmost column), to $28.652$ with $(\alpha,\beta,\varepsilon,k_0)=(1,0.1,1,20)$ (bottom row, leftmost column).

Equation~\eqref{eq-polysurv} provides our model a connection to a commonly used distribution in threshold-based methods for discrete extremes, specifically the integer generalised Pareto (IGP)
distribution \citep{Rohrbeck_2018} with conditional survival function: \[
\Pr(X> x|X> v) = \left(\frac{\xi(x-v)}{\sigma} + 1\right)^{-1/\xi},\qquad  x=v+1,v+2,\ldots
\] for \(v\in\mathbb Z^+, \sigma>0,\xi\in \mathbb R\), denoted as
\(X|X>v \sim  \mathrm {IGP}(\xi, \sigma, v)\) where \(\xi\) is the shape
parameter. As this is a regularly varying function with tail index $1/\xi$, the shape parameter $\xi$ is equivalent to the reciprocal of the tail index.

Using Equation~\eqref{eq-polysurv} and an approximation detailed in Appendix~\ref{sec-approx}, we can show that 
\begin{equation} \bar F(k|k>k_0)\approx\left(\frac{\beta(k-k_0)}{k_0^{\alpha}+\varepsilon+\beta} + 1\right)^{-\lambda^{*}/\beta}. \label{eq-approx} \end{equation}
As this means that the conditional limiting survival for $k>k_0$ can be approximated by the IGP distribution with $\xi=\beta/\lambda^{*}$, $\sigma=(k_0^\alpha + \varepsilon+\beta)/\lambda^*$, and $v=k_0$,
we note several points here. First, as the shape parameter $\xi=\beta/\lambda^{*}$ is the same as that of the reciprocal of the tail index derived in Section~\ref{sect.splice-gpa}, the tail heaviness is unaltered upon the approximation. Second, directly fitting the IGP distribution to the degrees, whilst seemingly appropriate, can only obtain estimates of $\xi$ and $\sigma$ but not identify the parameters \textit{of the spliced GPA model}. On the other hand, that the limiting survival in Equation~\eqref{eq-polysurv} is a function of the model parameters means that they can be inferred without incorporating the IGP distribution explicitly. This is supported by the absence of identifiability issues in the simulation study reported in Section~\ref{sec-sim}. Third, using the IGP distribution alone is not desirable as it requires choosing a threshold prior to inference, and fits to data subsetted by different thresholds are not directly comparable with each other as well as with the proposed model. Therefore, for a fair comparison, in our simulation studies and applications, we will consider the \textit{mixture} model by \cite{Lee24} that incorporates the IGP distribution for the tail, as an alternative to the proposed model.

\begin{figure}
\centering{
\includegraphics[width=\linewidth,height=\textheight,keepaspectratio]{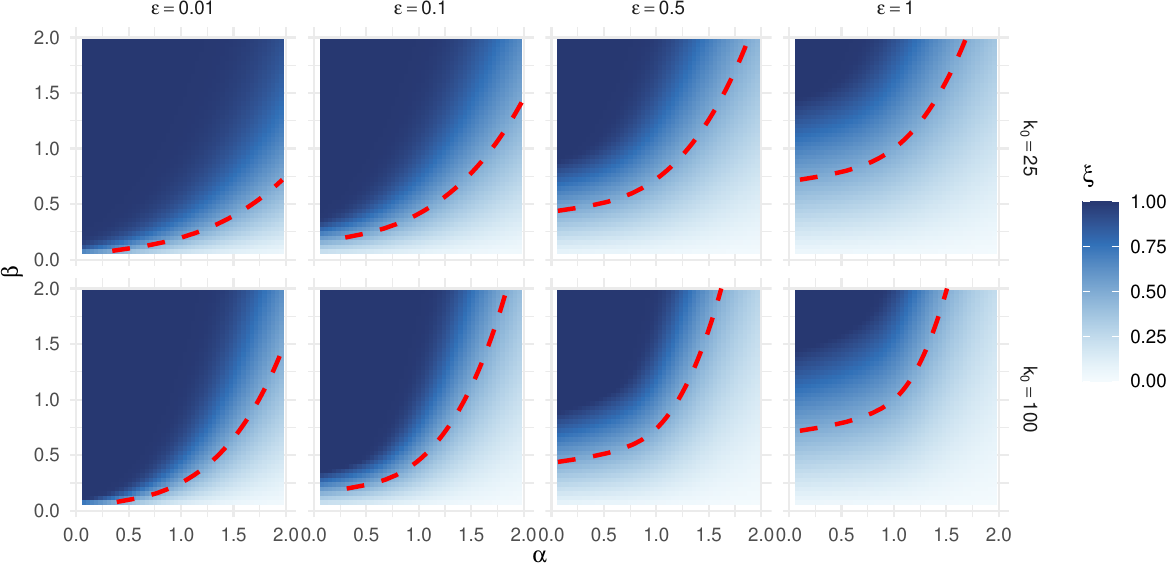}
}
\caption{\label{fig-polyheat}Heat maps of \(\xi=\beta/\lambda^{*}\) for various
combinations of the parameters of the proposed model. The red dashed line corresponds to $\xi=0.5$}
\end{figure}%

Lastly, without the IGP approximation, we consider the tail heaviness \textit{implied} by the parameters of the spliced GPA model, i.e. $\xi=\beta/\lambda^{*}$. The heat map of $\xi$ is shown in Figure~\ref{fig-polyheat} for various parameter
choices. The darker and lighter regions on the heat maps correspond to a
heavier and a lighter tail, respectively, and the red dashed line shows
combinations of \(\alpha\) and \(\beta\) that produce a limiting degree
distribution with the same tail index as the BA model.


\subsection{Inference}


As it is shown in Section~\ref{sec-numerical-properties} that the (limiting) degree distribution is an explicit function of the parameters of the spliced GPA model, statistical inference can be carried out as long as the degree sequence of the network snapshot is available. We outline the Bayesian inference procedure in this subsection.

For notational convenience, instead of denoting the degree of each node in the network, we consider the degree count vector \(\pmb n = (n_0, n_1, \ldots, n_M)\), where $n_i$ is the \textit{number of nodes} with degree $i$, and $M$ is the maximum degree. This means the total number of nodes is $\sum_{i=0}^Mn_i$ and the total degree is $\sum_{i=0}^Min_i$. As the large degrees are of more importance but usually with low counts, while the small degrees usually have much higher counts but might distort the fit, practically one might still truncate the data such that the minimum degree is $l\geq0$, and fit the model using the \textit{conditional} pmf derived from substituting Equation~\eqref{eq-pref} into Equation~\eqref{eq-pmf-origin}. The closed-form expression of the likelihood, as a function of the parameters, is:
\begin{equation}\phantomsection{
\begin{aligned}
L(\alpha,\epsilon,k_0,\beta|\pmb n, l) &=\prod_{i=l}^{k_0}\left[\frac{\lambda^{*}}{\lambda^{*}+i^{\alpha}+\varepsilon}\prod_{j=l}^{i-1}\frac{j^{\alpha}+\varepsilon}{\lambda^{*}+j^{\alpha}+\varepsilon}\right]^{n_i}\\
&\quad\times\left(\prod_{j=l}^{k_0-1}\frac{j^{\alpha}+\varepsilon}{\lambda^{*}+j^{\alpha}+\varepsilon}\right)^{\sum_{i=k_0+1}^{M}n_i}\\
&\quad\times\prod_{i=k_0+1}^{M}\left[\frac{\lambda^{*}\,\Gamma\left(\frac{k_0^{\alpha}+\varepsilon}{\beta}+i-k_0\right)\Gamma\left(\frac{\lambda^{*}+k_0^{\alpha}+\varepsilon}{\beta}\right)}{\beta\,\Gamma\left(\frac{\lambda^{*}+k_0^{\alpha}+\varepsilon}{\beta}+i-k_0+1\right)\Gamma\left(\frac{k_0^{\alpha}+\varepsilon}{\beta}\right)}\right]^{n_i}.
\end{aligned}
}\label{eq-likelihood}\end{equation} See Appendix~\ref{sec-likelihood} for the derivation.

As there have not been previous studies that used this particular preference function, there is no prior knowledge on the practical ranges of values for the parameters. Therefore, we assign to them uninformative priors (on the original scale of the parameters), with the specific values provided in Section~\ref{sec-sim}. While, for example, gamma priors can be used for $\alpha$, $\beta$ and $\epsilon$, care needs to be taken for the splice point $k_0$, which takes positive integer values. While $k_0=1$ reduces the model to a straight line and might therefore cause identifiability issues, a discrete uniform prior between $2$ and a large value e.g. $10^4$ has shown no issues in the simulation study as the posterior for $k_0$ rarely concentrates on small values.

As the likelihood is a closed-form expression of a small number of parameters, we can carry out inference using the Metropolis-Hastings algorithm in Markov chain Monte Carlo (MCMC) to draw samples from the posterior. For $\alpha$, $\beta$, and $\epsilon$, which are updated simultaneously, a trivariate Gaussian distribution, with the current values as the mean, is used as the proposal distribution. Similarly but separately, for $k_0$, we use a random walk with equal probability for positive and negative direction, and a step size drawn from a Poisson distribution with mean 1. The usual steps of carrying out MCMC are taken, including discarding an initial portion of the samples as the burn-in period, and checking the convergence through diagnostics such as the effective sample sizes and the trace plots.

\section{Simulations}\label{sec-sim}

This section aims to show that the parameters of the spliced model
(and preference function) can
be recovered from the degrees of a network originating from the spliced GPA model. To demonstrate this, we simulate networks from the model with several parameter combinations and use the likelihood from Equation~\eqref{eq-likelihood} with a set of uninformative priors to perform Bayesian inference. Our results show that posterior means accurately recover the true parameter values with narrowing credible intervals.

The set up for fitting the spliced GPA model is the same throughout this section. An adaptive Metropolis-Hastings MCMC algorithm was used for 100,000 iterations, with the first 50,000 discarded as burn-in, and the following priors:
\begin{align*}
    \alpha &\sim \text{Gamma(1, 0.001)},\\
    \varepsilon &\sim \text{Gamma(1, 0.001)},\\
    \beta &\sim \text{Gamma(1, 0.001)},\\
    k_0 &\sim U(2, 10^4),
\end{align*}
 where $\text{Gamma}(a, b)$ denotes the Gamma distribution having mean $a/b$, and $U(a,b)$ is the discrete uniform distribution supported on $\{a, a+1, \ldots, b-1, b\}$.
  
\subsection{One-shot experiment}
First, we present a one-shot experiment for three different scenarios (shown in Table~\ref{tab:oneshot}) showing the posteriors for the parameters, preference function, and survival in each case. We also include a comparison to the IGP mixture model from \cite{Lee24} with survival:

$$
\bar F(k) = \begin{cases}
    [1-\phi_u]\frac{\sum_{i=k+1}^u (i+1)^a \theta^{i+1}}{\sum_{i=0}^u (i+1)^a \theta^{i+1}},&k=0,1,\ldots,u,\\
    \phi_u\left(1 + \frac{\xi(k-u)}{\xi u +\sigma}\right)^{-1/\xi}, &k=u+1,u+2,\ldots,
\end{cases}
$$
where $u$ is a given threshold and $a\in\mathbb R$ and $\theta\in(0,1]$ are the parameters of the truncated Zipf-polylog distribution, and $\xi\in\mathbb R, \sigma>0$ are the parameters of the IGP distribution, and $\phi_u\in(0,1)$. The model is fitted and the parameters ($a,\theta, \xi, \sigma,u$) estimated using the R package \verb|crandep| \citep{crandep} available on CRAN. This allows us to compare the spliced GPA model and the IGP mixture model when it comes to modelling the full set of degrees, and the tail by comparing the shape parameter $\xi$ to the spliced GPA equivalent $\beta/\lambda^*$. We choose to compare to this model as it is also a two component spliced mixture aimed at modelling network degrees and is therefore a direct competitor to the spliced GPA model.

\begin{table}[ht]
    \centering
    \begin{tabular}{c|c|c|c|c|c}
        \hline
        Scenario & Power ($\alpha$) & Zero appeal ($\varepsilon$) & Slope ($\beta$) & Splice point ($k_0$) &  Size ($N$)\\ 
        \hline
          I & 0.5 & 0.1 & 0.5 & 10 & $10^5$ \\
          II & 1 & 0.5 & 0.1 & 10 & $10^5$\\
          III & 1.5 & 1 & 1 & 10 & $10^5$
    \end{tabular}
    \caption{Parameter combinations of selected scenarios of the simulations}
    \label{tab:oneshot}
\end{table}

\begin{figure}[ht]
    \centering
    \begin{tabular}{ccc}
    (I)&(II)&(III)\\
       \includegraphics[width=0.3\linewidth]{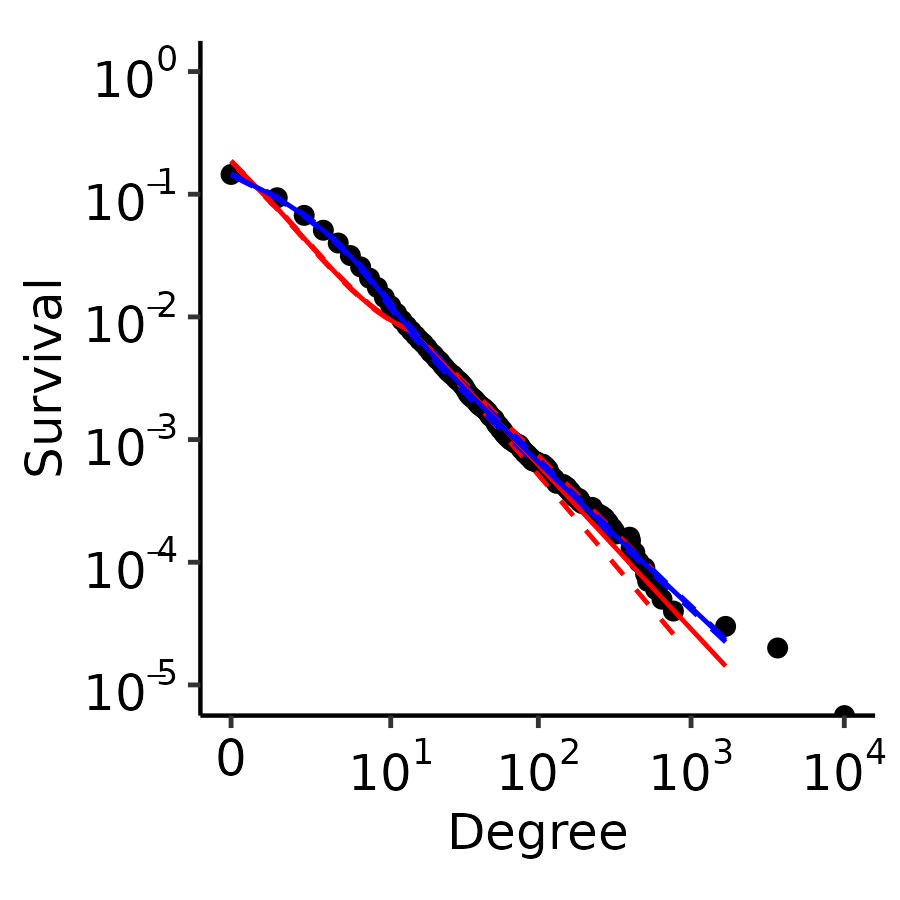}   &  \includegraphics[width=0.3\linewidth]{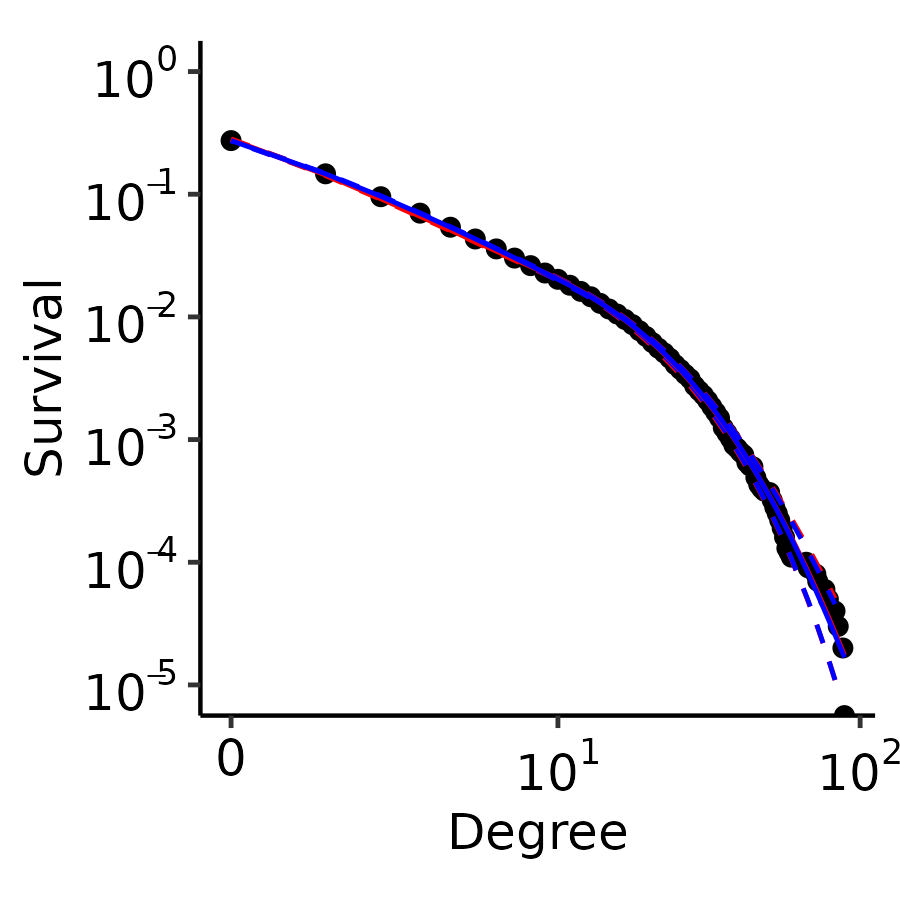}&\includegraphics[width=0.3\linewidth]{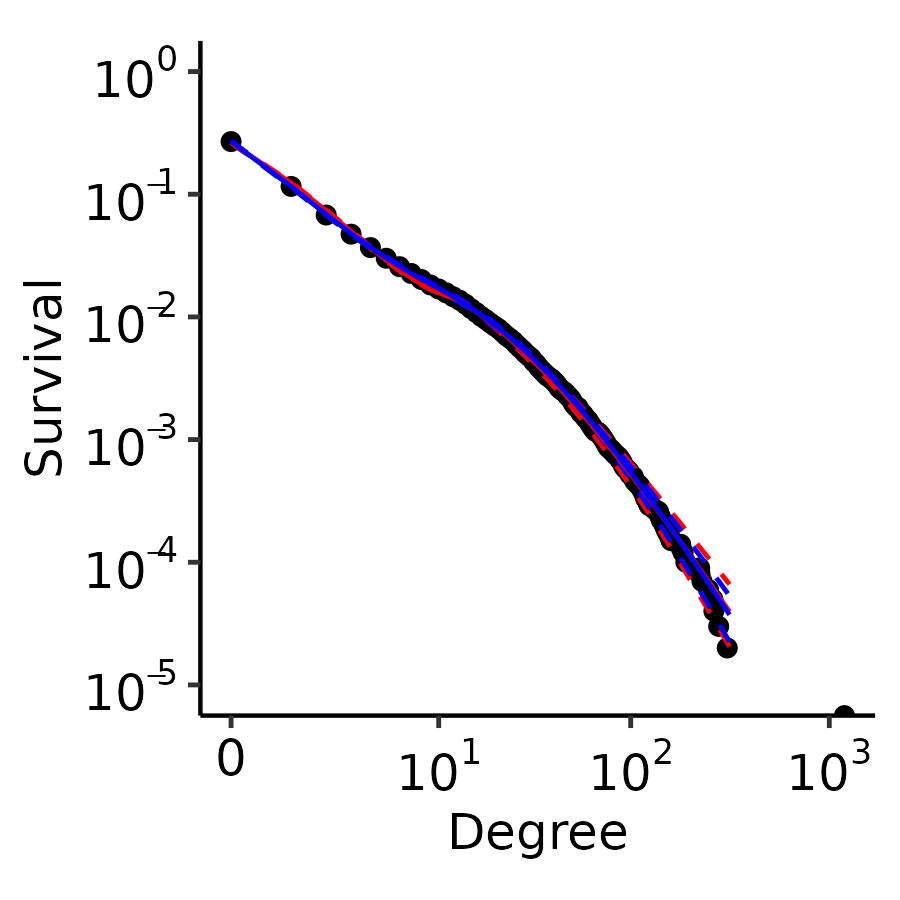} \\
    \end{tabular}

    \caption{Empirical (black dots), fitted spliced GPA model (blue line), fitted IGP mixture model (red line) survivals for networks simulated under the scenarios in Table~\ref{tab:oneshot}}
    \label{fig:oneshotsurvs}
\end{figure}
\begin{figure}[ht]
    \centering
    \begin{tabular}{ccc}
    (I)&(II)&(III)\\
         \includegraphics[width=0.3\linewidth]{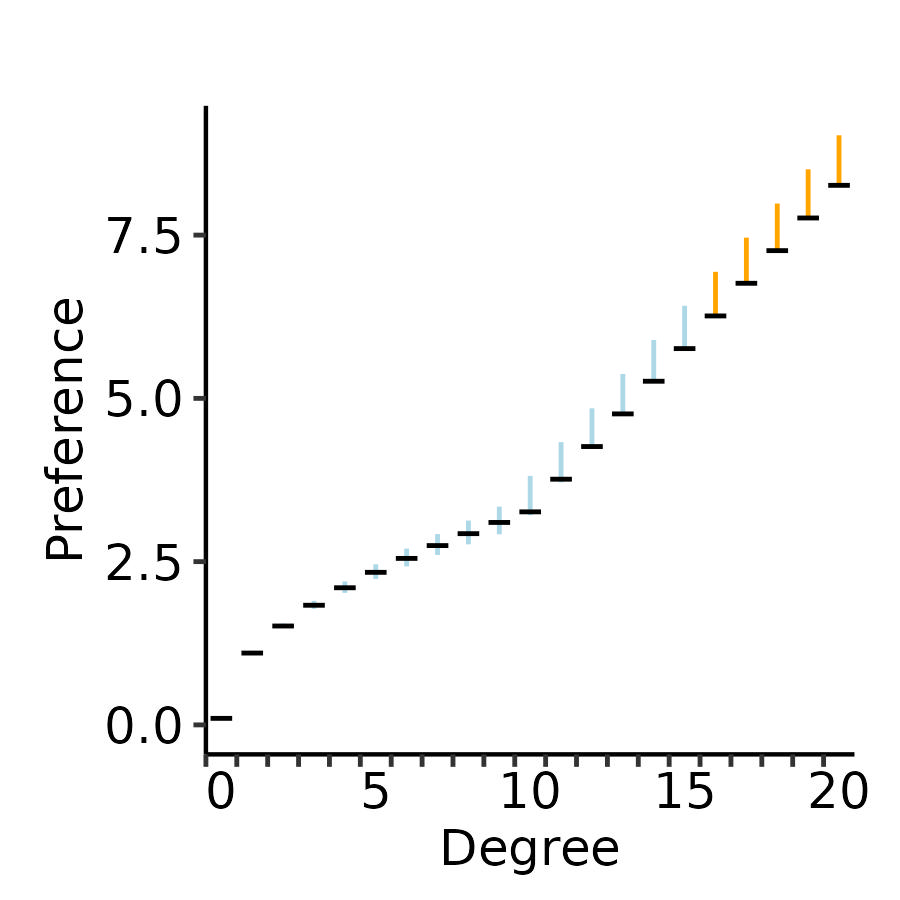} &
         \includegraphics[width=0.3\linewidth]{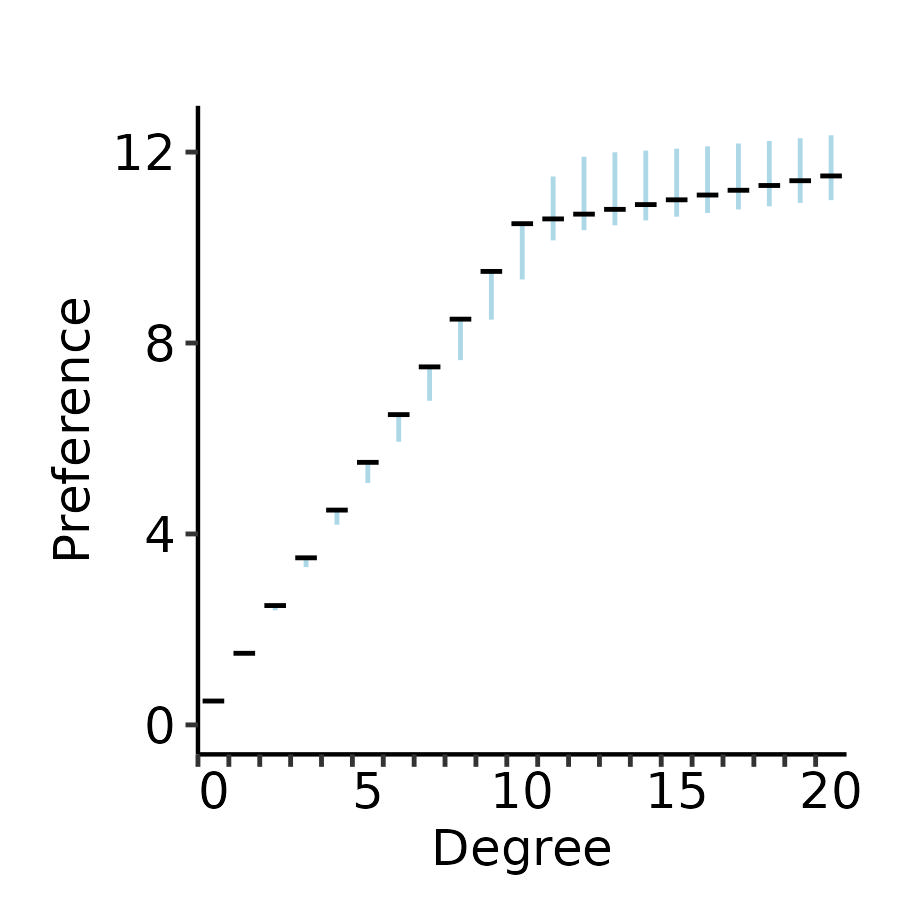} &
         \includegraphics[width=0.3\linewidth]{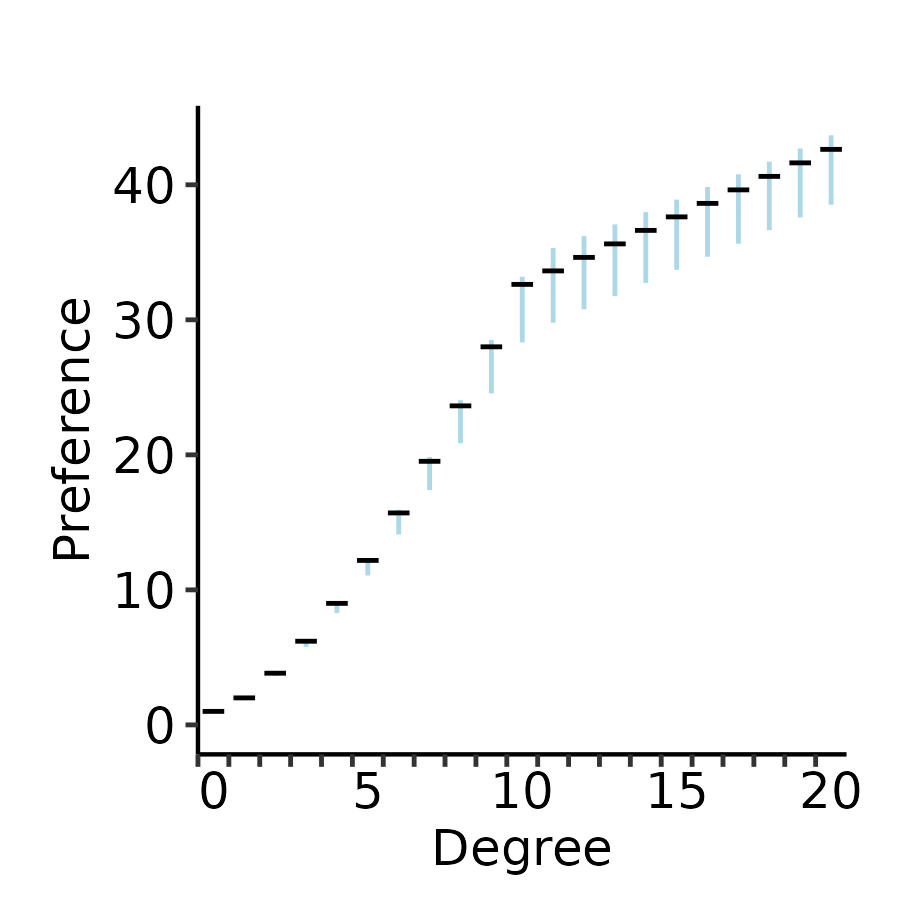}
    \end{tabular}
    \caption{True preference function (black horizontal bars) with credible intervals (vertical bars) of preference function from fitted spliced GPA model for networks simulated under the scenarios in Table~\ref{tab:oneshot}}
    \label{fig:oneshotprefs}
\end{figure}

As shown in Figure~\ref{fig:oneshotsurvs}, the fitted survival functions of our proposed model are comparable to existing bulk-tail approaches such as the IGP mixture model, across the scenarios in Table~\ref{tab:oneshot}. However, our framework offers a distinct advantage: as illustrated in Figure~\ref{fig:oneshotprefs}, the parameters of the preference function $b(k)$ can be successfully recovered from snapshot degree data. To the best of our knowledge, our model is the first to achieve this, making the estimated preference function a novel feature that lacks a direct counterpart among competing methods, giving additional information about the network's dynamic growth mechanism.

Aside from the three previous scenarios, we also performed the one-shot experiment for a total of 36 parameter combinations and two different network sizes. Specifically, the scenarios consisted of the following combinations:
\begin{align*}
    \alpha~(\text{Power})& \in \{   0.5,1,1.5\},\\
    \varepsilon~(\text{Zero appeal}) &\in \{0.1,0.5,1\},\\
    \beta~(\text{Slope}) &\in \{0.1,0.5,1,1.5\},\\
    k_0~(\text{Splice point}) &= 10,\\
    N (\text{Size}) &\in\{10^4, 10^5\}.
\end{align*}

\begin{figure}[!ht]
\centering{
\begin{tabular}{c c}
   \includegraphics[width=0.41\linewidth,height=\textheight,keepaspectratio]{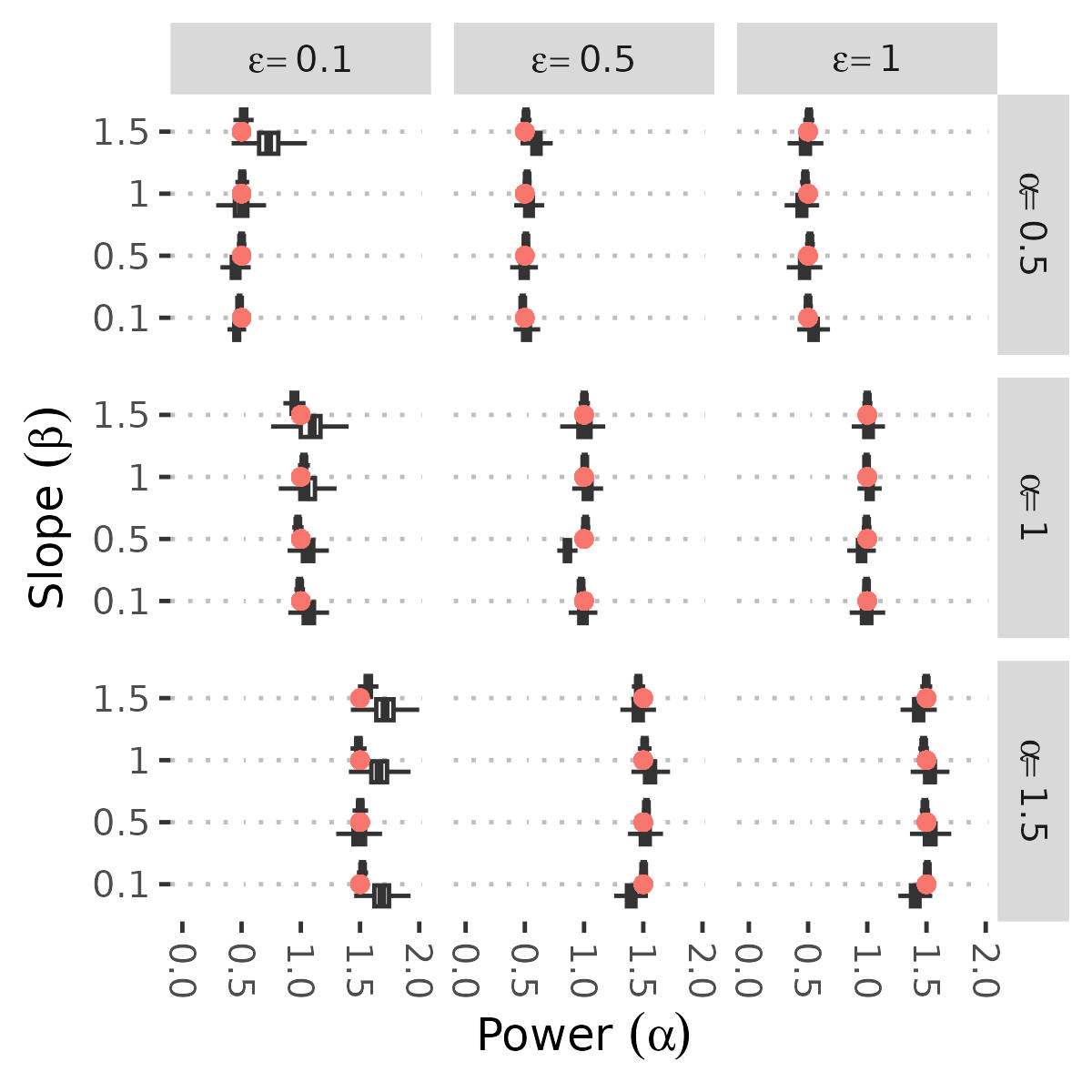}  & \includegraphics[width=0.41\linewidth,height=\textheight,keepaspectratio]{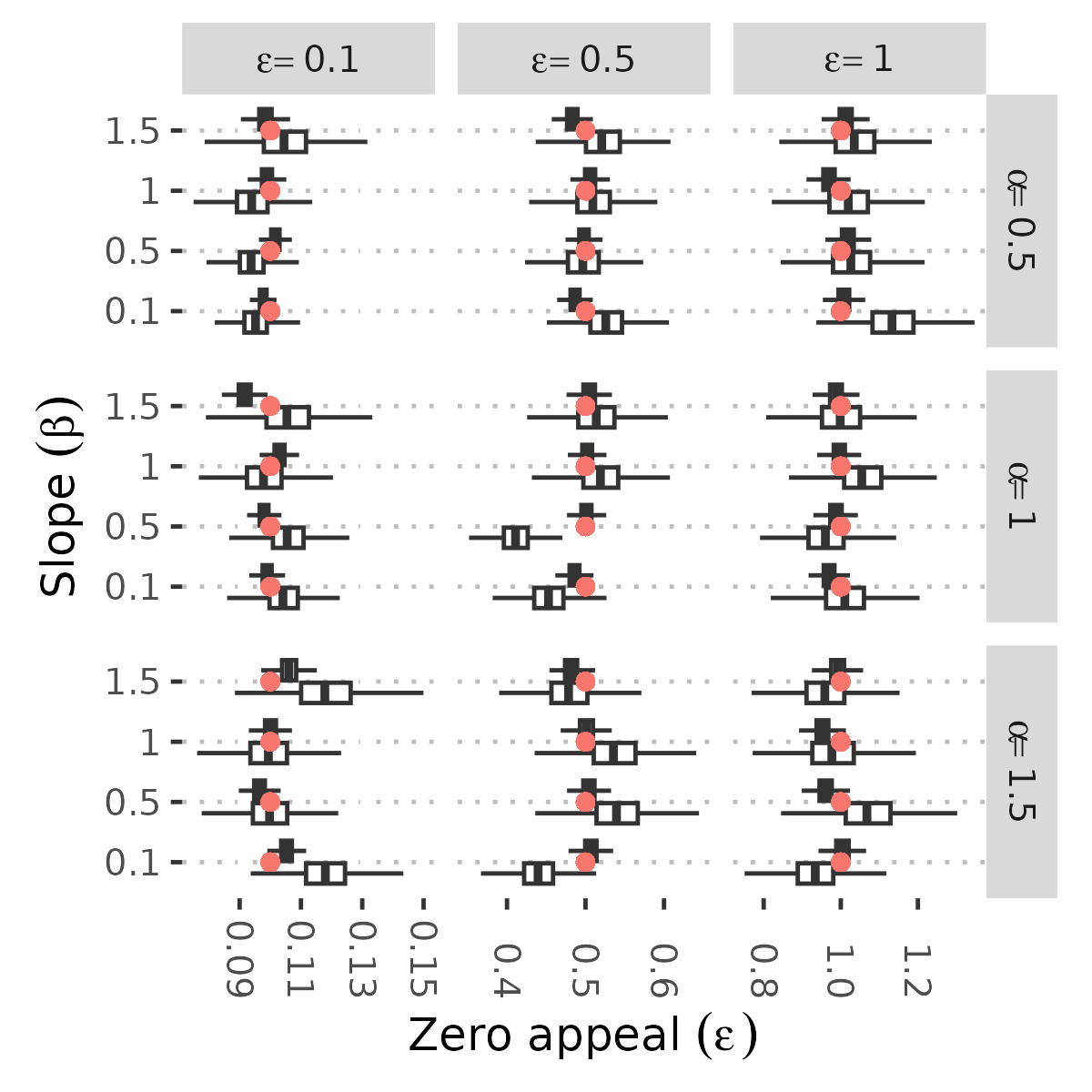} \\
    \includegraphics[width=0.41\linewidth,height=\textheight,keepaspectratio]{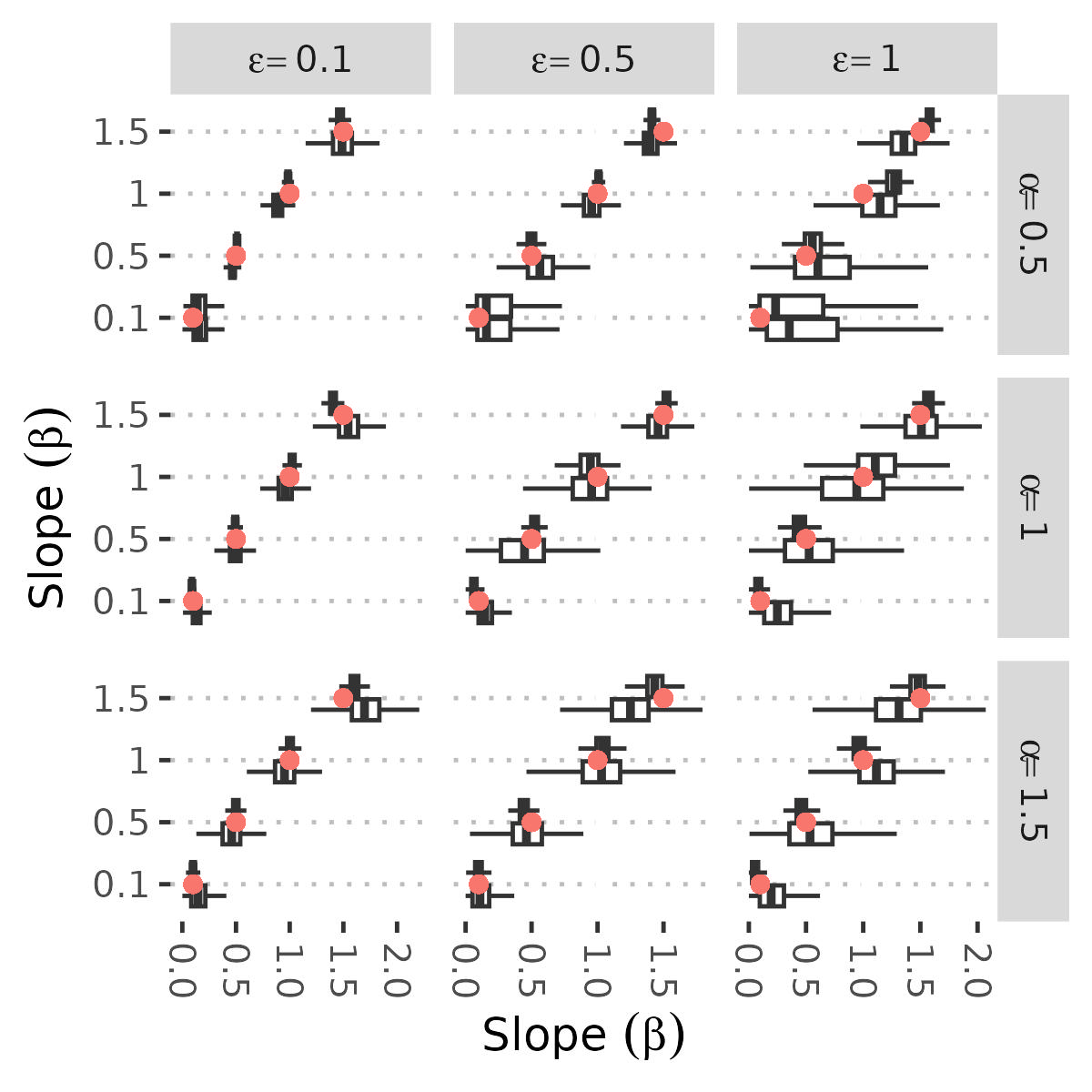} & \includegraphics[width=0.41\linewidth,height=\textheight,keepaspectratio]{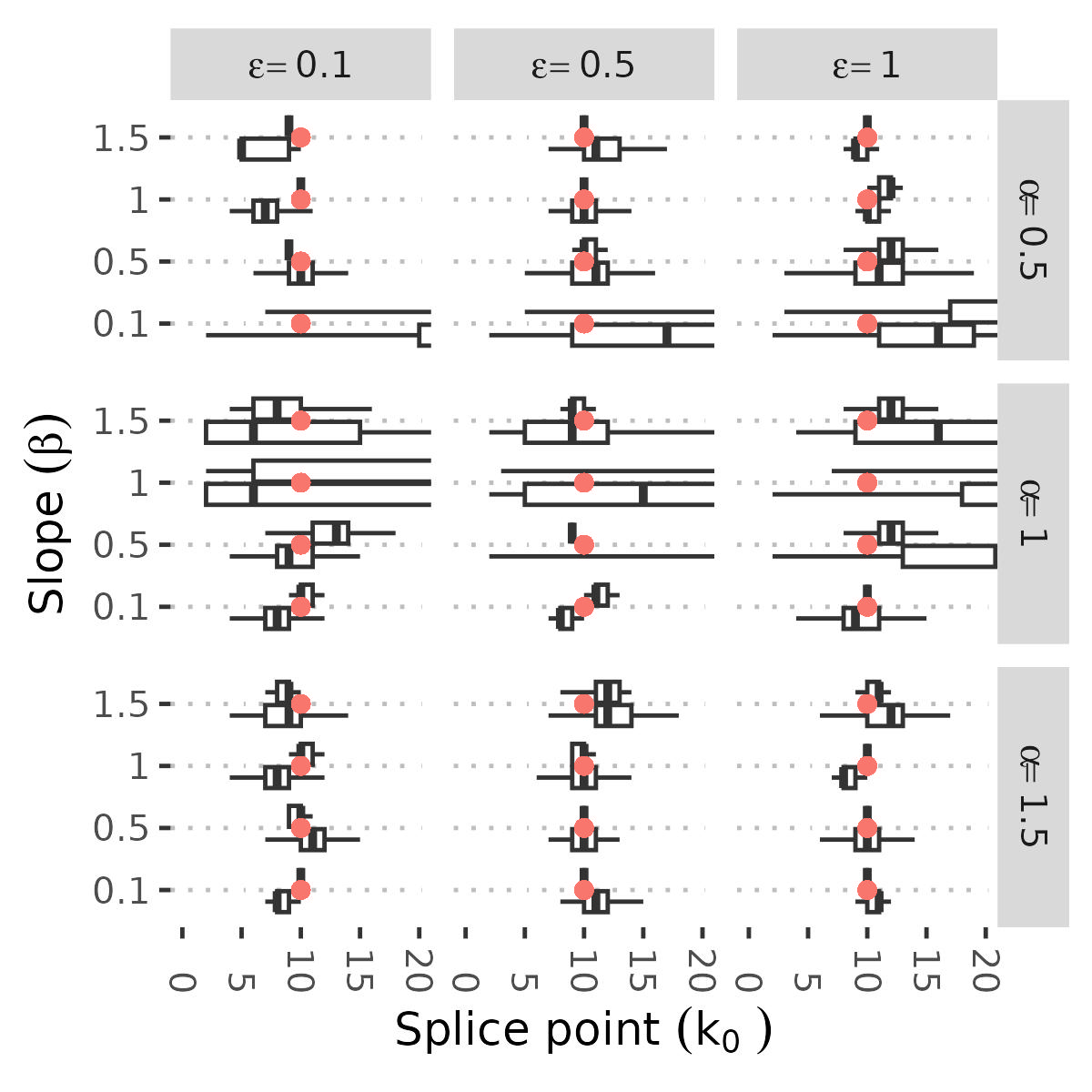}
\end{tabular}
}

\caption{\label{fig-oneshotfull10} Posterior boxplots of parameters for data
simulated from the spliced GPA model with various combinations of
(\(\alpha\),\(\beta\),\(\varepsilon\)), \(k_0=10\), and $N=10^5$ (top row) and $N=10^4$ (bottom row). The orange dots are the true parameter values. Some boxplots of $k_0$ extend beyond the readable range}

\centering{
\begin{tabular}{c c}
   \includegraphics[width=0.42\linewidth,height=\textheight,keepaspectratio]{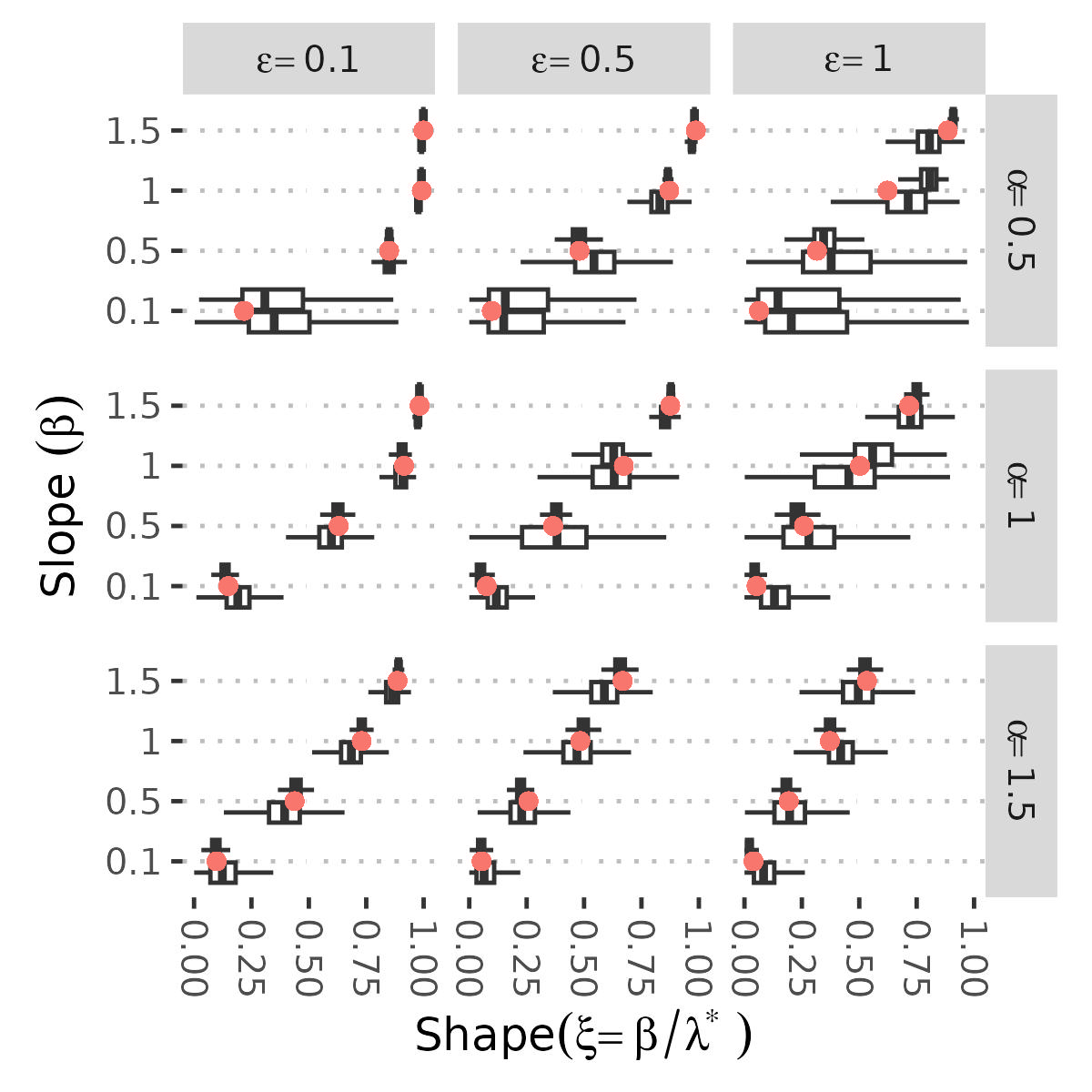}  & \includegraphics[width=0.42\linewidth,height=\textheight,keepaspectratio]{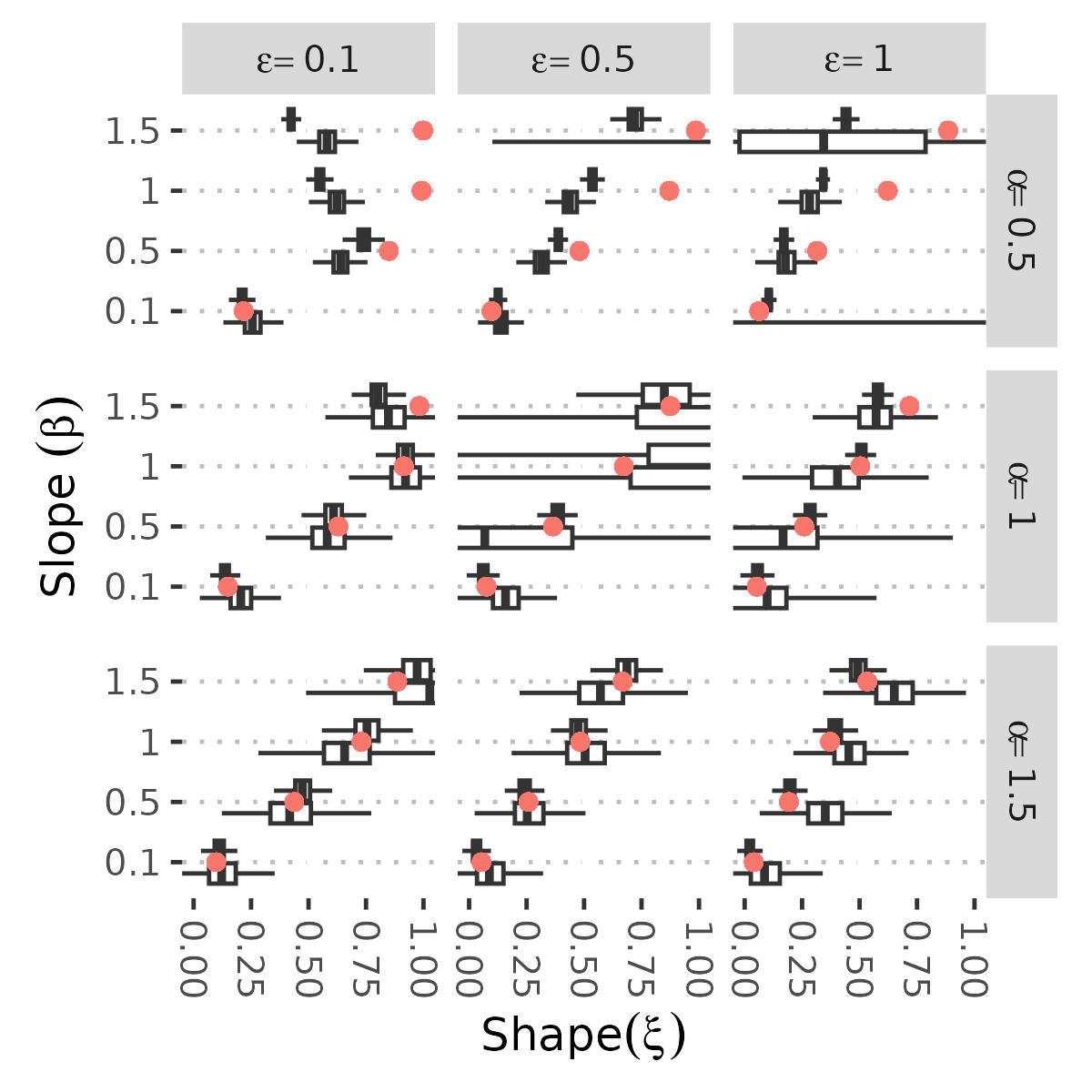} 
\end{tabular}
}

\caption{\label{fig-oneshotfullshape} Posterior boxplots of shape parameter $\xi$ in IGP mixture model (right) and equivalent in spliced GPA model (left) for data
simulated from the spliced GPA model with various combinations of
(\(\alpha\),\(\beta\),\(\varepsilon\)), \(k_0=10\), and $N=10^5$ (top row) and $N=10^4$ (bottom row). The orange dots are the true parameter values}
\end{figure}%

The results of this expanded one-shot experiment are shown in Figure~\ref{fig-oneshotfull10}, which shows that the parameters are generally well recovered with the parameters used to generate a network of $10^5$ nodes being recovered more reliably than that of a network with $10^4$ nodes. The cases where the model is not performing well correspond to either when the maximum degree is not much higher than the splice point, meaning there is little to no data for inferring the splice point and slope, or when the slope is close to (or exactly) tangent with the polynomial part of the preference function at the splice point, making it indiscernible over the range of data simulated. 

Figure~\ref{fig-oneshotfullshape} focuses on the estimation of $\xi$, the reciprocal of the tail index, comparing the estimates obtained from the spliced GPA model and the IGP mixture model. As it is the key parameter governing the limiting tail behaviour of the degree distribution, its precise estimation is crucial. As shown in Figure~\ref{fig-oneshotfullshape}, the spliced GPA model outperforms the IGP mixture model for smaller values of $\alpha$ and yields comparable estimation performance for larger values of $\alpha$.

\subsection{Monte Carlo Simulation Study}
\begin{figure}[!htb]
    \centering
    \begin{tabular}{ccc}
            (I)&(II)&(III)\\
         \includegraphics[width=0.3\linewidth]{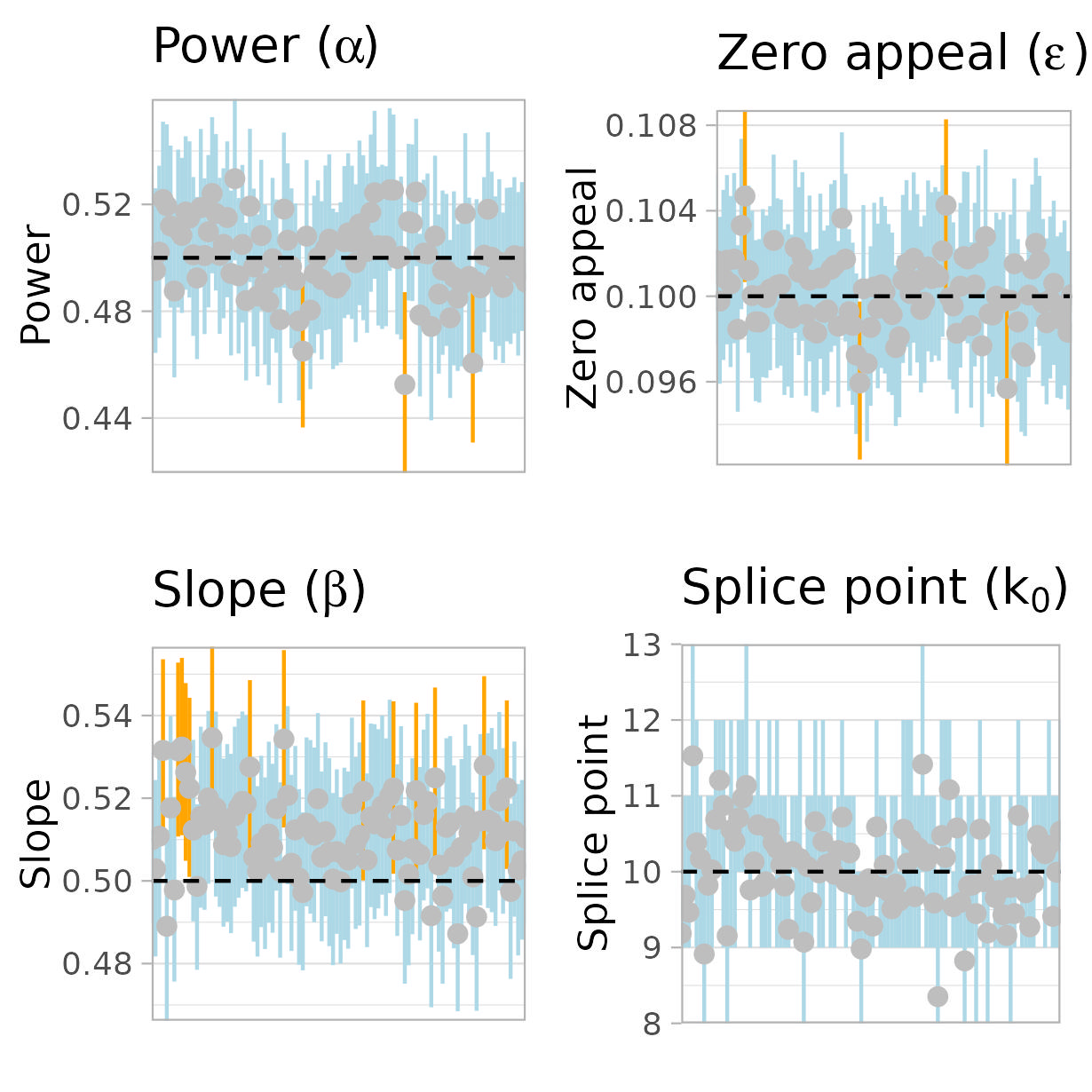} &
         \includegraphics[width=0.3\linewidth]{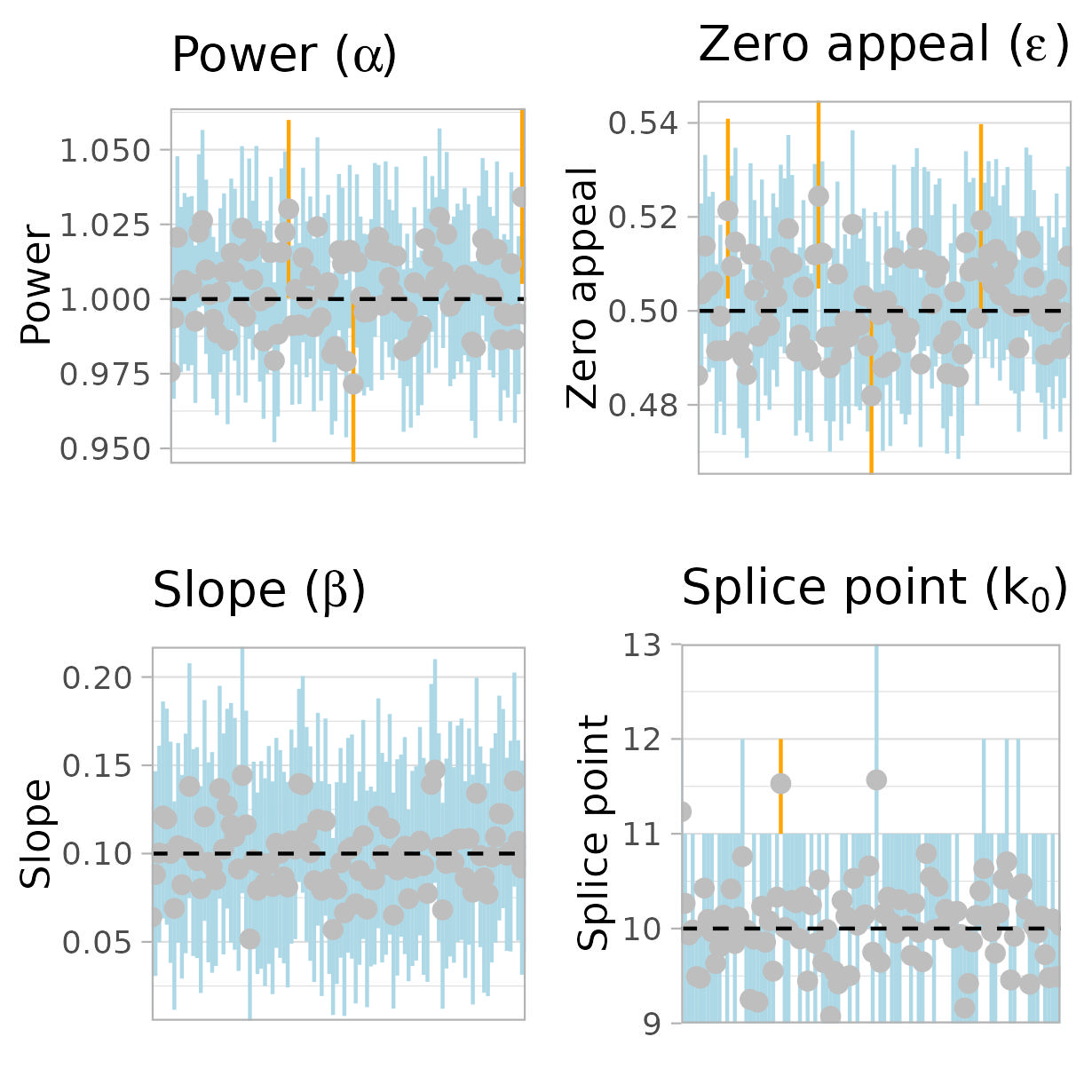} &
         \includegraphics[width=0.3\linewidth]{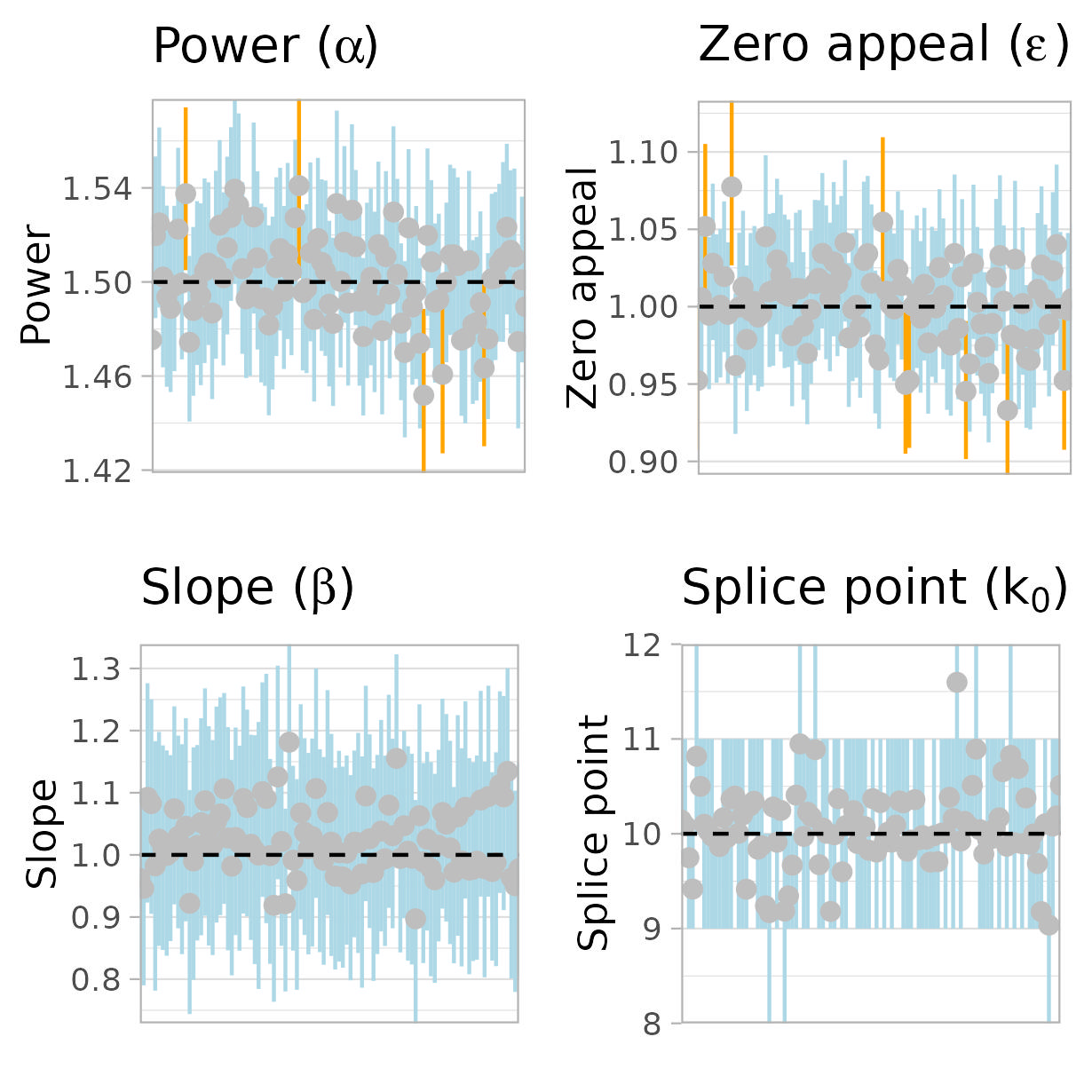}
    \end{tabular}
    \caption{Posterior means (grey 
    dots) and credible intervals (vertical bars) of parameters for 100 replications of simulating and fitting spliced GPA model in scenarios from Table~\ref{tab:oneshot}. The dashed lines are the true parameter values}
    \label{fig:MC_params}
\end{figure}

\begin{figure}[!htb]
    \centering
    \begin{tabular}{ccc}
        (I) & (II) & (III)\\
         \includegraphics[width=0.3\linewidth]{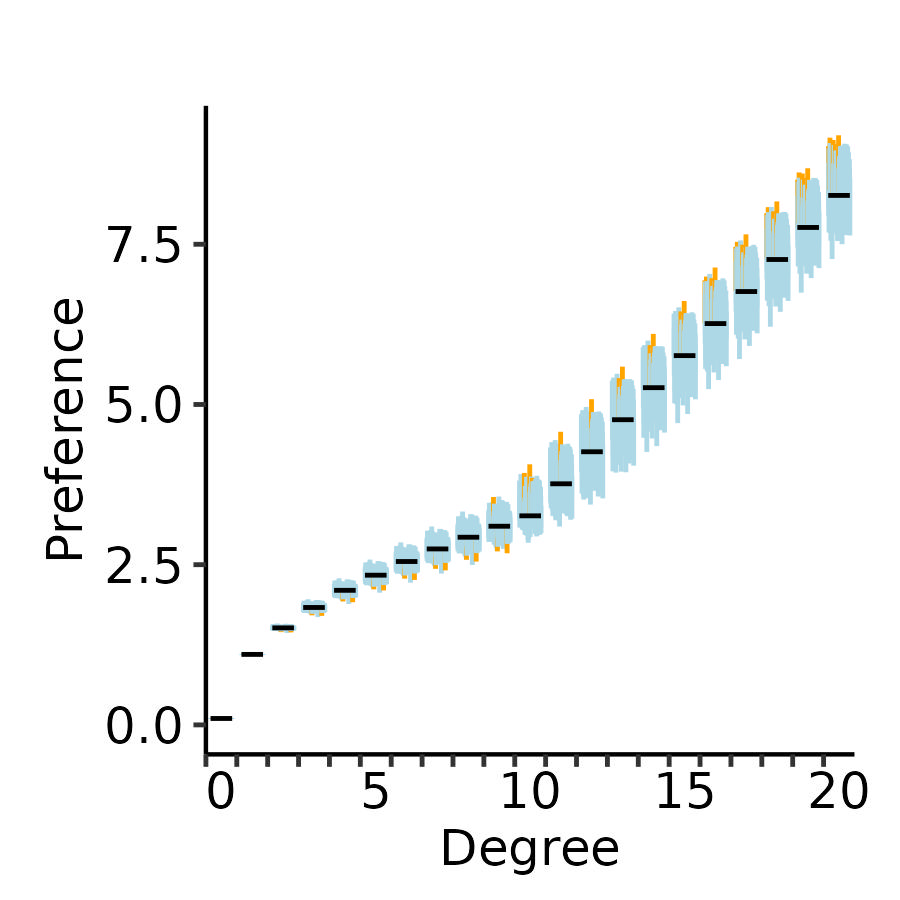} &
         \includegraphics[width=0.3\linewidth]{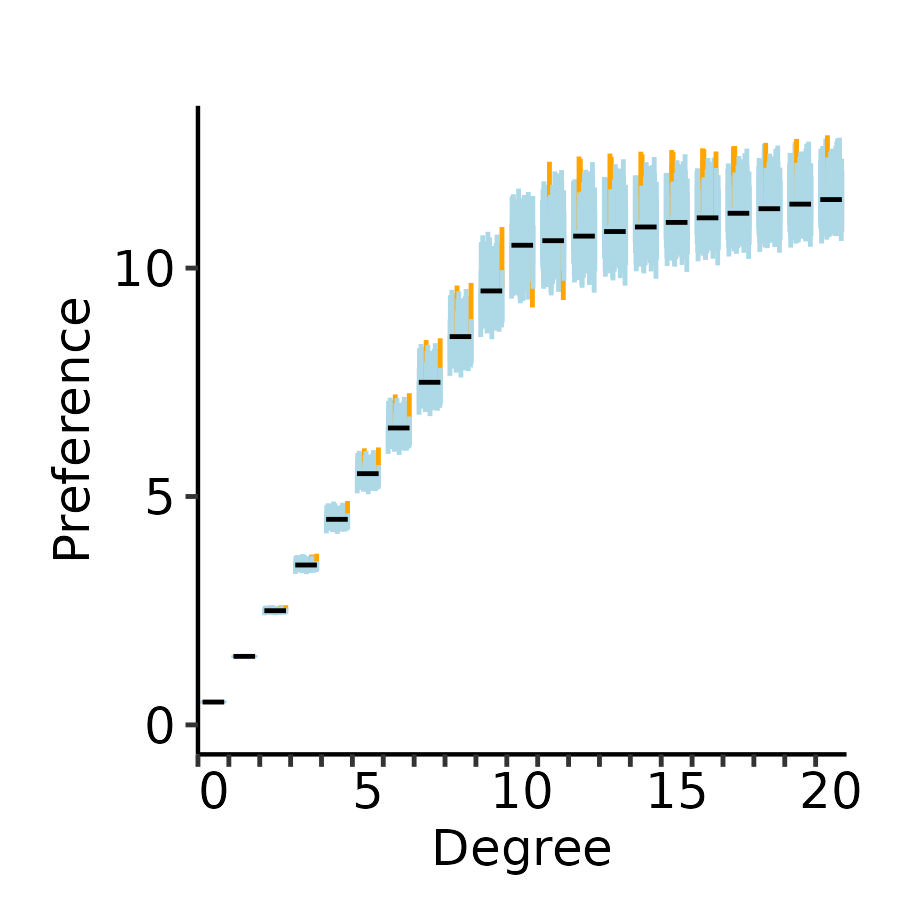} &
         \includegraphics[width=0.3\linewidth]{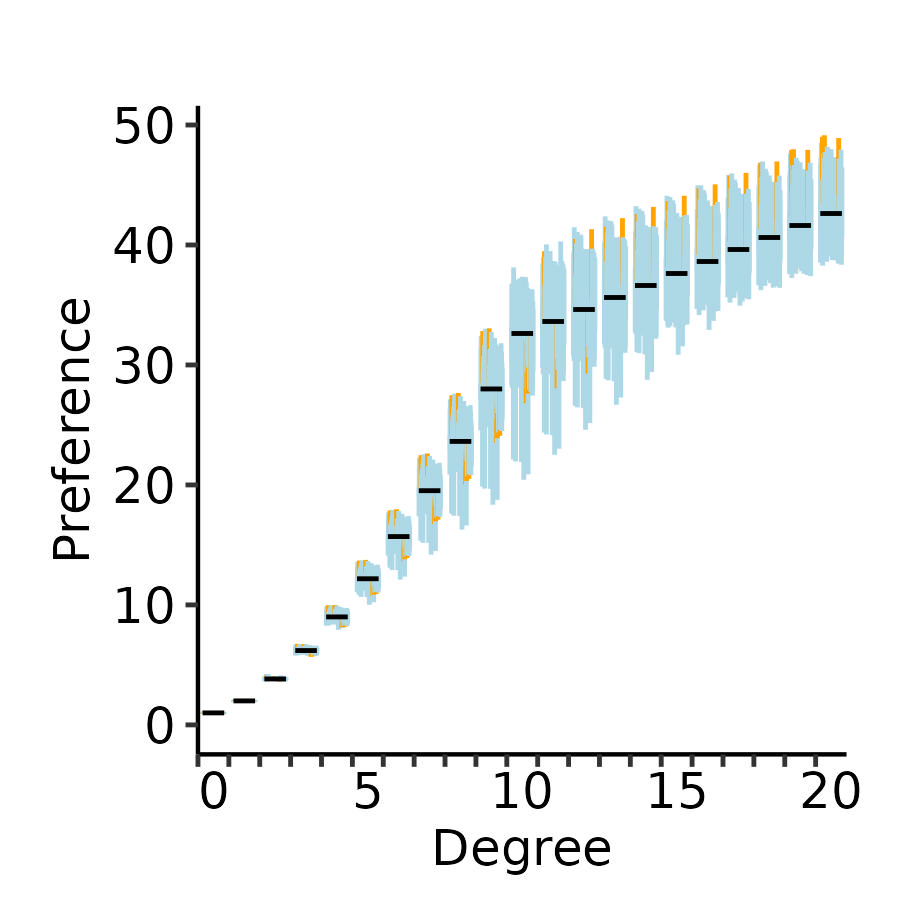}
    \end{tabular}
    \caption{True preference function (black horizontal bars) with credible intervals (vertical bars) of preference function from 100 replications of fitting spliced GPA model to networks simulated under scenarios in Table~\ref{tab:oneshot}}
    \label{fig:MC_pref}
\end{figure}

To assess the reliability of fitting the spliced GPA model we repeated all of the previous experiments 100 times.
Figure~\ref{fig:MC_params} shows that the model parameters are consistently recovered across the three scenarios, while Figure~\ref{fig:MC_pref}, which serves as the Monte Carlo analogue of Figure~\ref{fig:oneshotprefs}, demonstrates that the estimated preference function $b(k)$ closely matches the true underlying growth mechanism across all three scenarios.

\begin{figure}[!htb]
\centering{
\begin{tabular}{c c }
   \includegraphics[width=0.45\linewidth,height=\textheight,keepaspectratio]{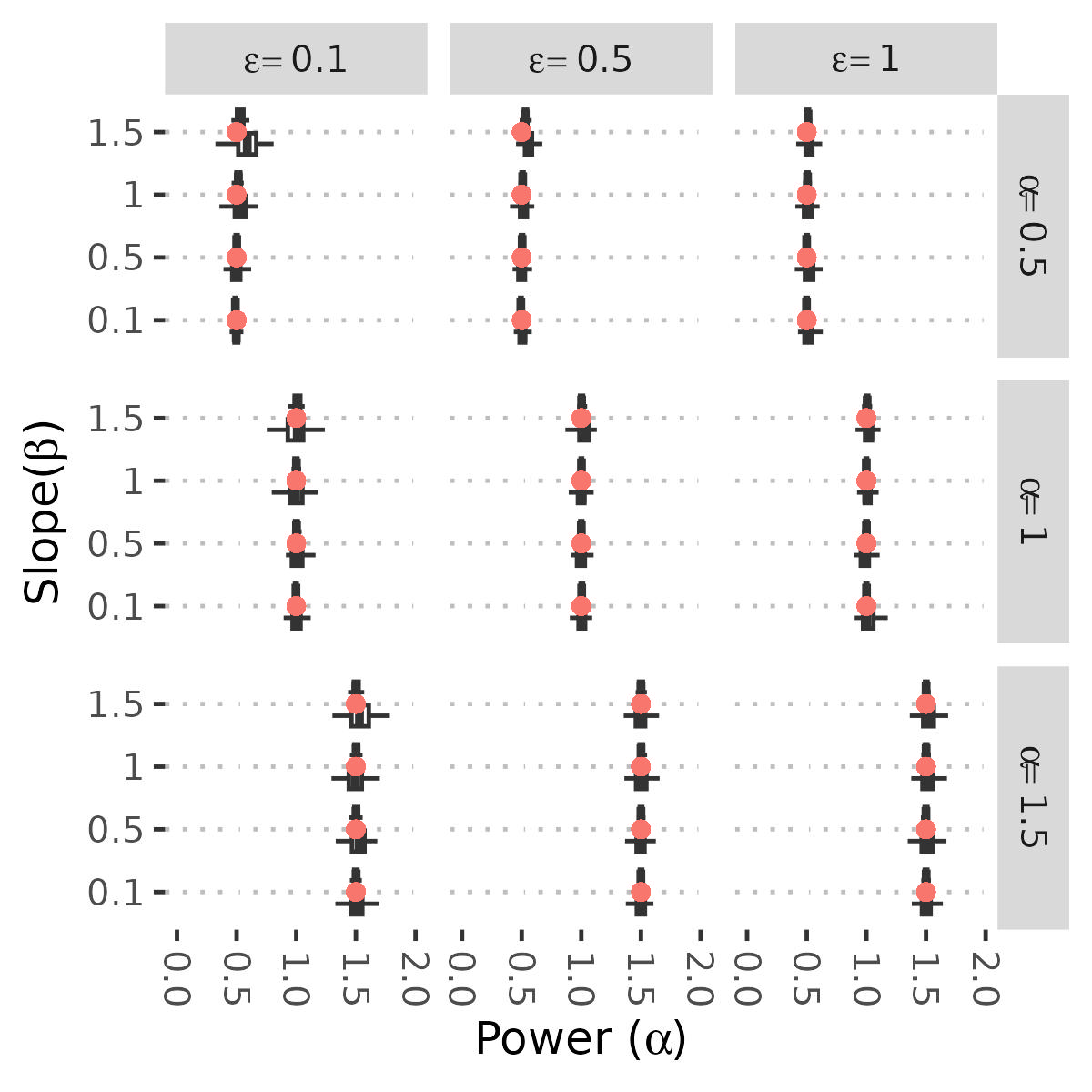}  & \includegraphics[width=0.45\linewidth,height=\textheight,keepaspectratio]{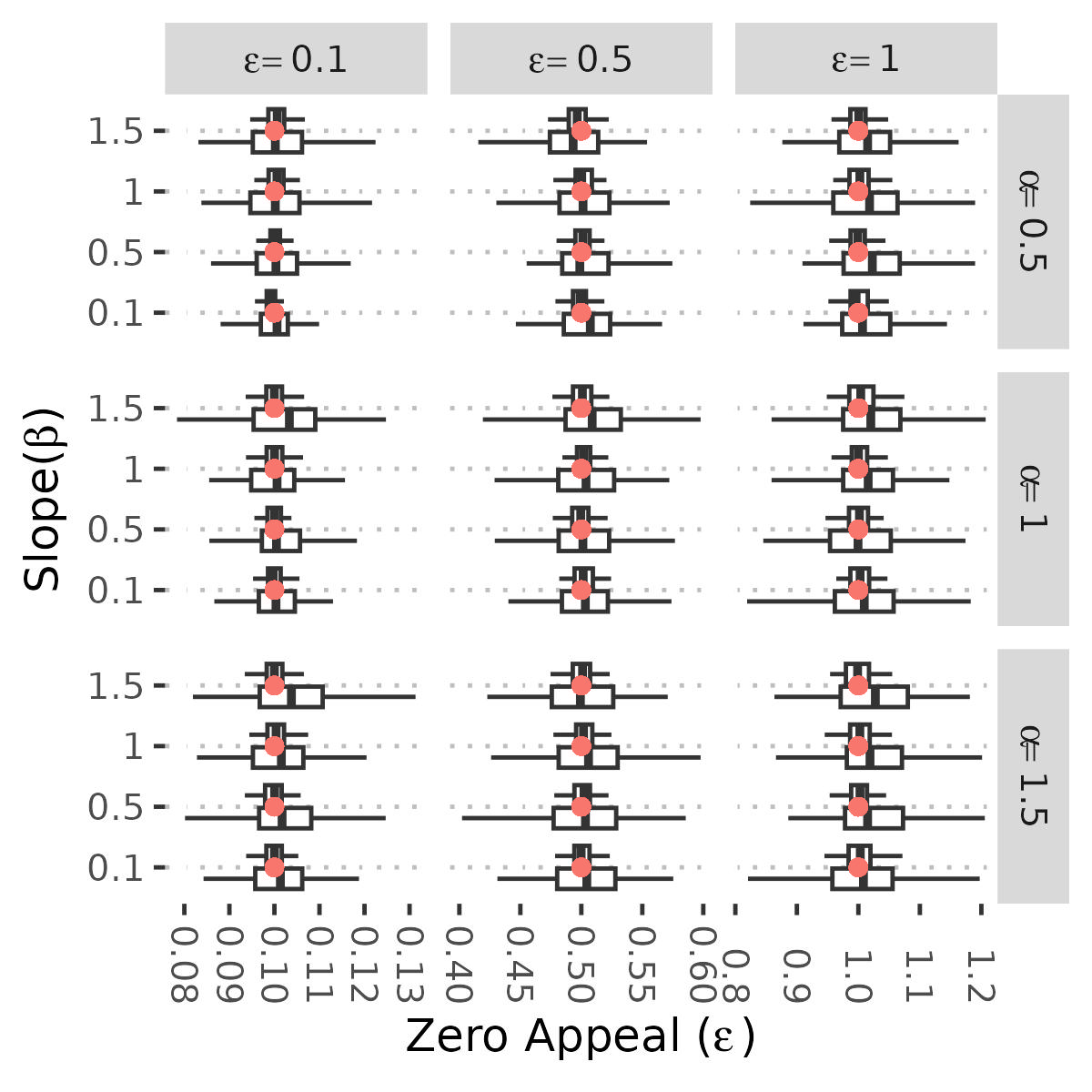} \\
    \includegraphics[width=0.45\linewidth,height=\textheight,keepaspectratio]{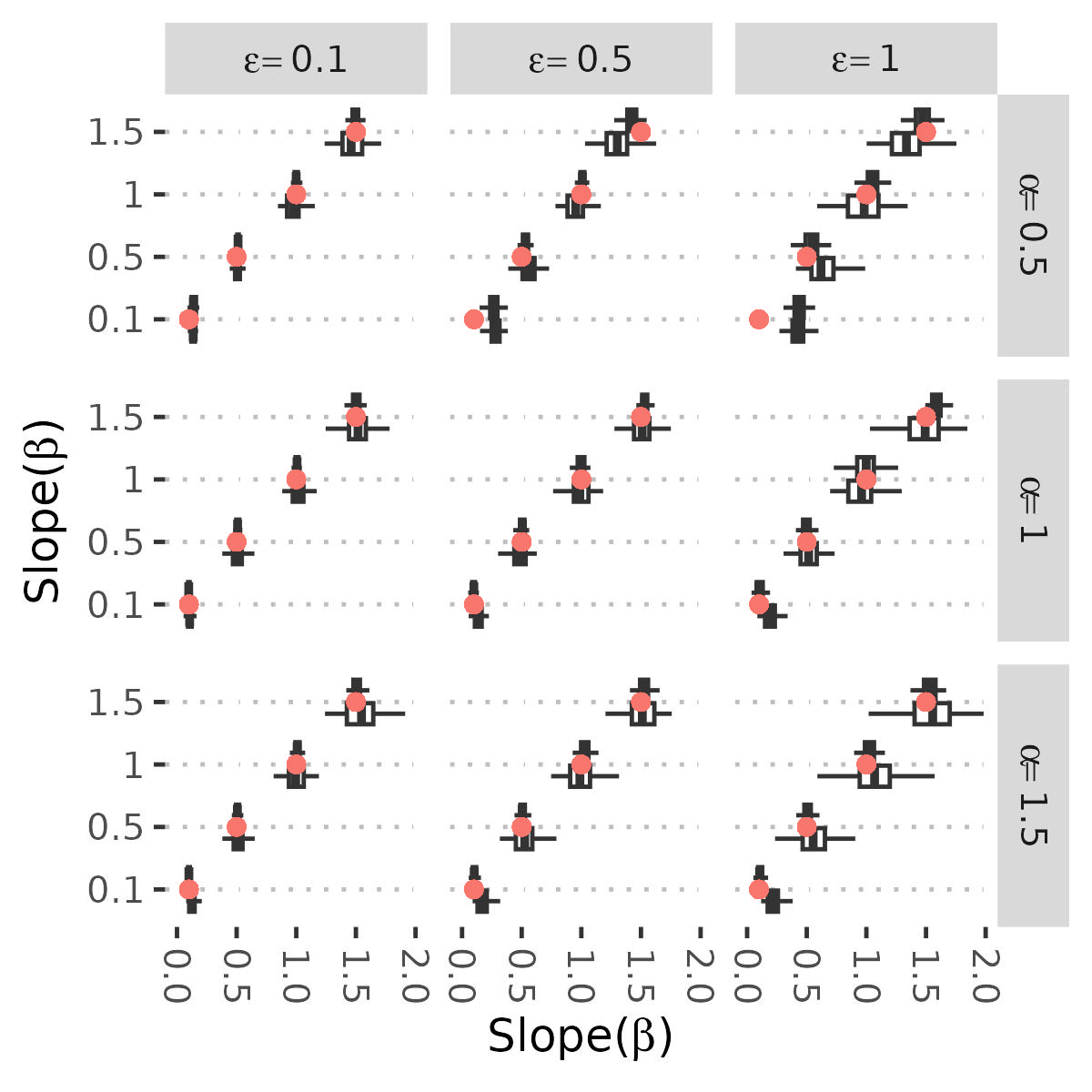} & \includegraphics[width=0.45\linewidth,height=\textheight,keepaspectratio]{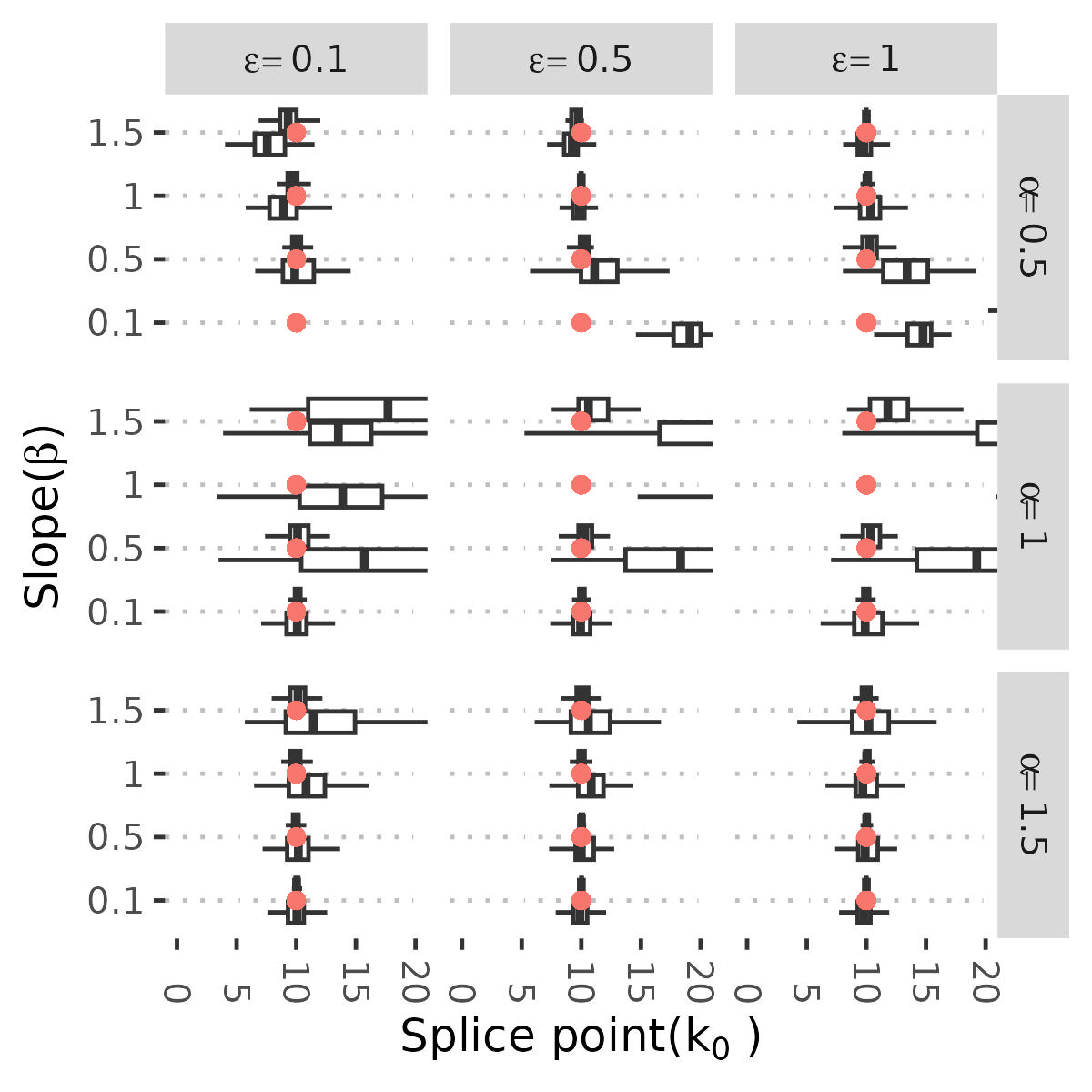}
\end{tabular}
}

\caption{\label{fig:MCfullpars}: Boxplot of posterior means of parameters for data
simulated from the proposed model with various combinations of
(\(\alpha\),\(\beta\),\(\varepsilon\)), \(k_0=10\), and $N=10^5$ (top row) and $N=10^4$ (bottom row). The orange dots are the true parameter values}
\end{figure}%

Finally, we expanded the study across the full set of parameter combinations. Figure~\ref{fig:MCfullpars} presents the Monte Carlo analogue of Figure~\ref{fig-oneshotfull10}, displaying boxplots of the posterior means across all $100$ replications for each scenario. Across the parameter space, the posterior means remain tightly concentrated around their true values, with minor estimation difficulties occurring only in the edge cases previously identified in the single run experiment where the splice point $k_0$ is challenging to identify. We also see the increase in accuracy gained as the network size $N$ increases. For a complete analysis across different sample and network sizes, detailed estimation tables and root mean square errors for simulation $(M=30,60, \text{and } 100)$ and network sizes $(N=10^4 \text{ and } 10^5)$ are provided in Section~2 of the Supplementary Material. As expected, increasing either the simulation size $M$ or the network size $N$ improves estimation precision and reduces errors, but the model already gives reliable parameter estimates even at $M=30$.

This section demonstrates that the parameters and hence the generating mechanism can be inferred from a snapshot of the network degrees alone. The competitive predictive performance of our spliced GPA model motivates the application to real world network data in Section~\ref{sec-real}.

\section{Applications to Real Data} \label{sec-real}

In this subsection, we fit the proposed model to the degree
distributions of various real networks and learn about the mechanics of
their growth. While we also compare the fit to that of the mixture
distribution by \citet{Lee24} we note that the proposed method has the
additional benefit of learning directly about the growth of a network
from the inference results. The data consists of 12 networks sourced
from \href{konect.cc}{KONECT} and the
\href{https://networkrepository.com}{Network Data Repository} \citep{nr}:

\begin{itemize}
\tightlist
\item
  \texttt{as-caida20071105}: network of autonomous systems of the
  Internet connected with each other from the CAIDA project
\item
  \texttt{dimacs10-astro-ph} : co-authorship network from the
  ``astrophysics'' section (astro-ph) of arXiv
\item
  \texttt{ego-twitter}: network of twitter followers
\item
  \texttt{facebook-wosn-wall}: subset of network of Facebook wall posts
\item
  \texttt{maayan-faa}: USA FAA (Federal Aviation Administration)
  preferred routes as recommended by the NFDC (National Flight Data
  Center)
\item
  \texttt{maayan-Stelzl}: network representing interacting pairs of
  proteins in humans
\item
  \texttt{moreno-blogs-blogs}: network of URLs found on the first pages
  of individual blogs
\item
  \texttt{opsahl-openflights}: network containing flights between
  airports of the world.
\item
  \texttt{pajek-erdos}:
  \(\text{co-authorship network around Paul Erd\H{o}s}\)
\item
  \texttt{reactome}: network of protein--protein interactions in humans
\item
  \texttt{sx-mathoverflow}: interactions from the StackExchange site
  \href{https://mathoverflow.net/}{MathOverflow}
\item
  \texttt{topology}: network of connections between autonomous systems
  of the Internet
\end{itemize}

The parameter inferences that follow were obtained using the same priors
and MCMC sampling method as detailed in Section~\ref{sec-sim}.

\begin{figure}
\centering{
\includegraphics[width=115mm,height=\textheight,keepaspectratio]{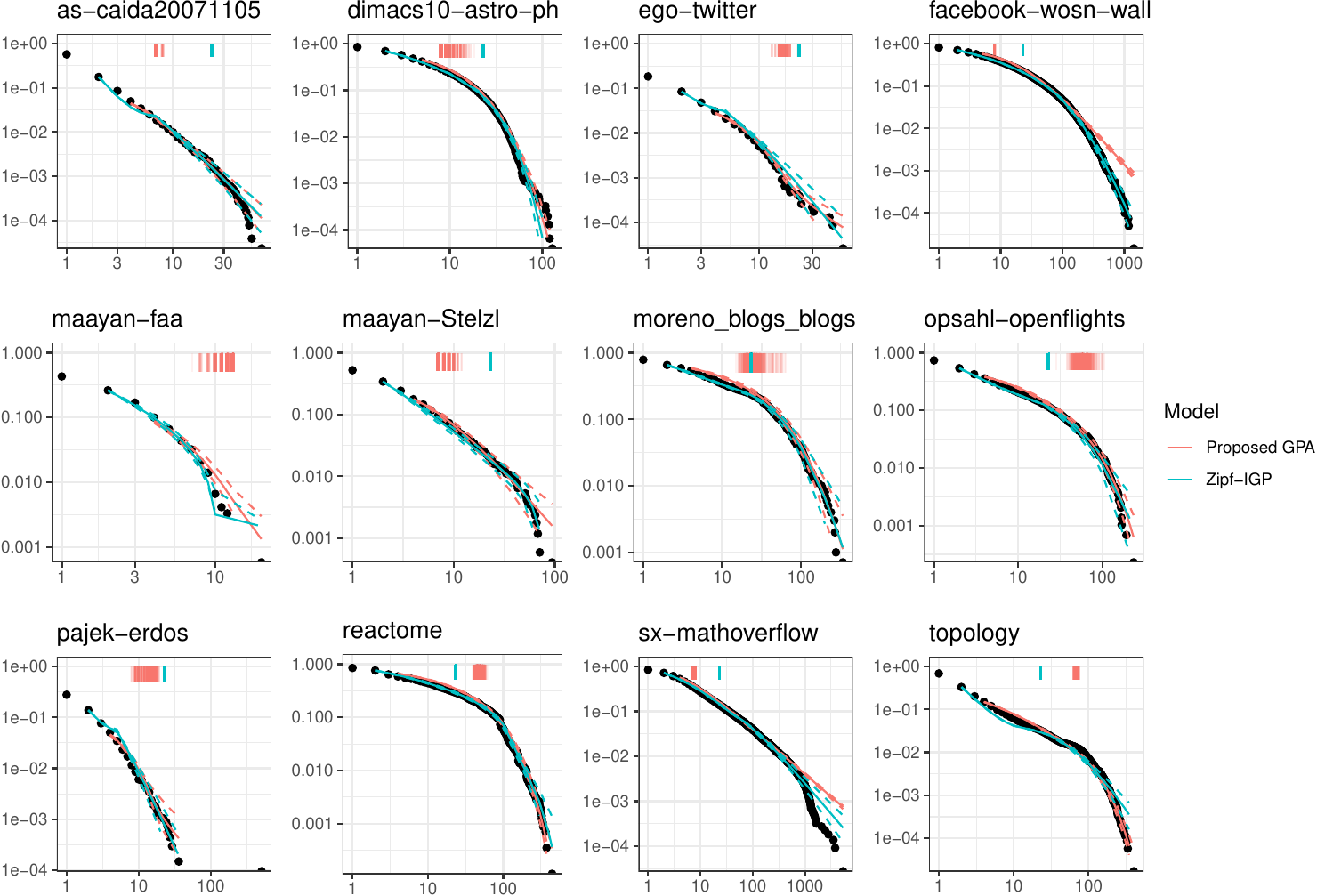}
}
\caption{\label{fig-real1}Empirical (black dots) and posterior medians
of the fitted survival function of the spliced GPA model (solid red)
for several real data sets and their 95\% credible intervals (dotted
red), the posterior medians (solid blue) and 95\%
credible intervals (dotted blue) for the IGP mixture model. The vertical lines at the top are the corresponding \(k_0\) and threshold}
\centering{
\includegraphics[width=115mm,height=\textheight,keepaspectratio]{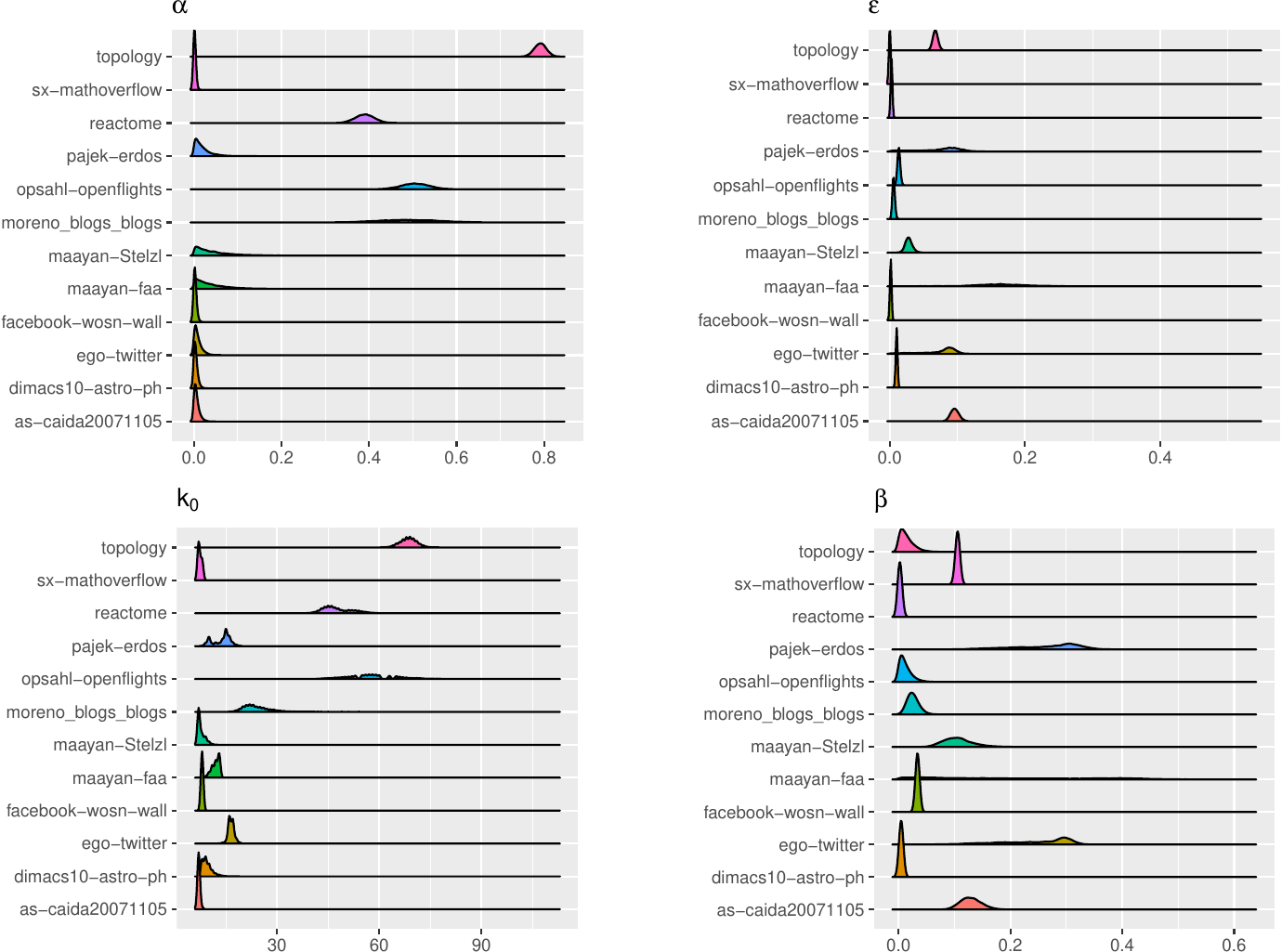}
}
\caption{\label{fig-parplot}Posterior densities of the parameters of the
spliced GPA model fitted to real data}
\end{figure}%

Figure~\ref{fig-real1} displays the posterior estimates of the survival
function for various data sets, obtained from fitting the spliced GPA
model and the IGP mixture model. While the two models give a similar fits in most cases, where the spliced GPA model fits well we gain additional information
about the preference function, assuming that the network did evolve
according to it. The corresponding posterior densities for the parameters are shown in Figure~\ref{fig-parplot} where wide ranges of value are observed across all parameters.

\begin{figure}[!ht]
\centering{
\includegraphics[width=119mm,height=\textheight,keepaspectratio]{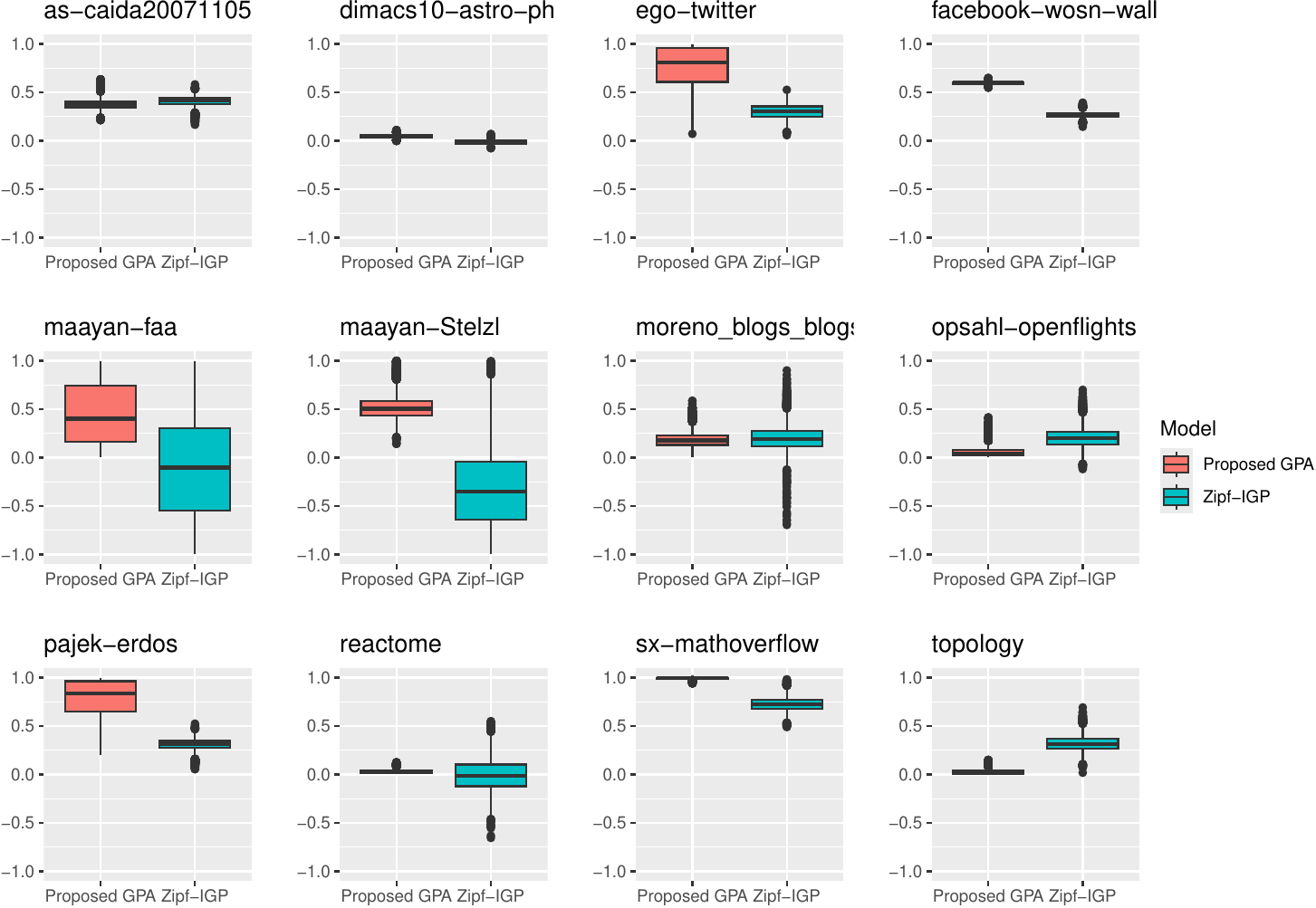}
}
\caption{\label{fig-shapes}Posterior estimates of shape parameter of
IGP mixture (right) and the equivalent under the spliced GPA model (left)}
\centering{
\includegraphics[width=119mm,height=\textheight,keepaspectratio]{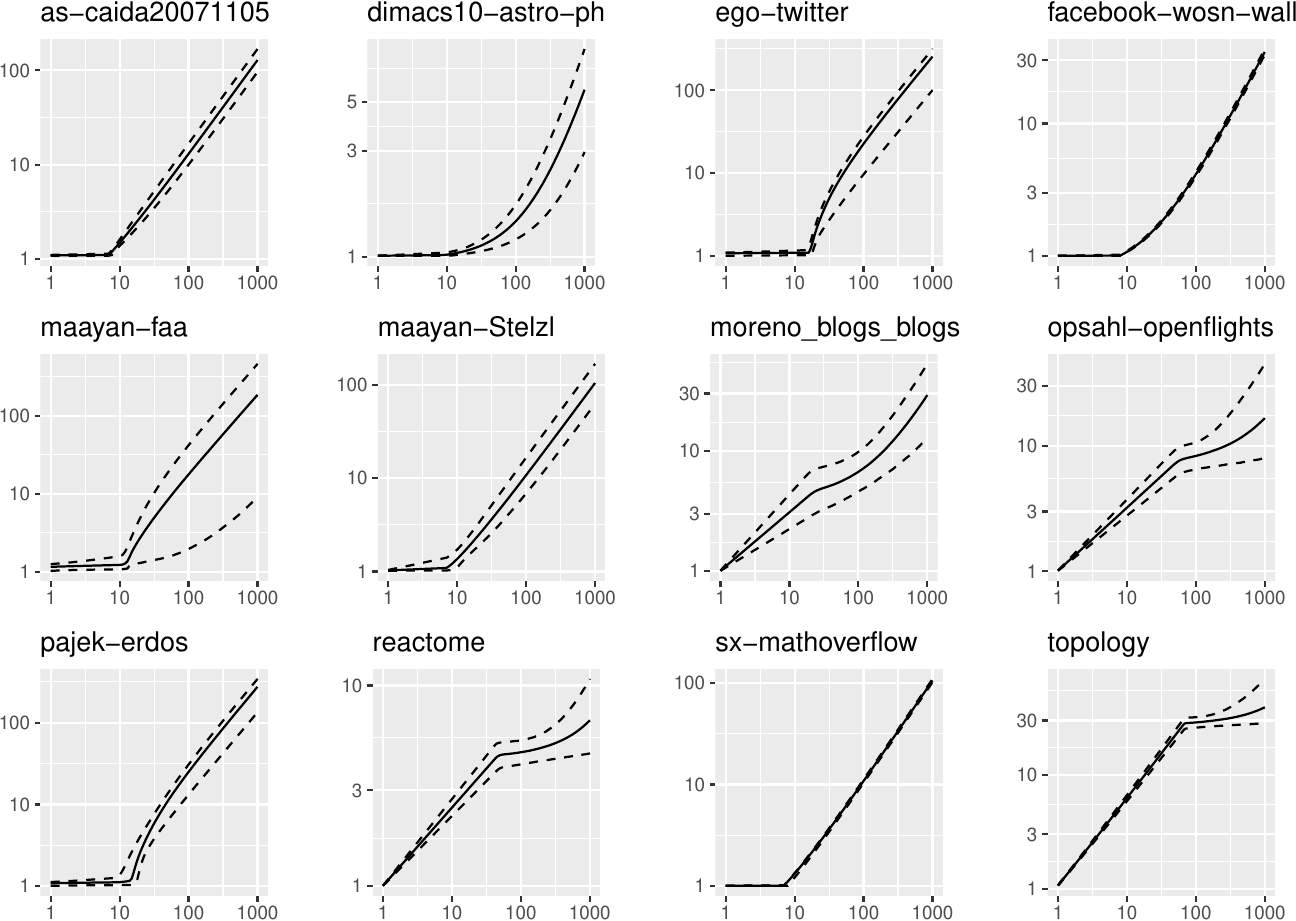}
}
\caption{\label{fig-pa}Posterior median for the preference function
(solid) with 95\% credible interval (dashed) on log-log scale}
\end{figure}

Figure~\ref{fig-shapes} shows the posterior of the shape parameter
\(\xi\) obtained from the IGP mixture model alongside the posterior of the
equivalent shape parameter \(\beta/\lambda^*\) obtained from fitting the
proposed model. Generally, the proposed model performs similarly
to the IGP mixture when estimating the tail behaviour of the degree
distribution. In the cases of substantial discrepancies, it is either
because the proposed model fits the tail better than the IGP mixture
model does, or because of the threshold being too low, forcing almost
all of the data to be modelled by the linear part of the GPA. This again
shows the effects that small degrees have on this model.

Figure~\ref{fig-pa} shows the estimated preference function \(b(k)\)
alongside the 95\% credible interval on a log-log scale. Although the
credible interval becomes very large for the largest degrees, this is
expected as not all of these networks had data in that region, and for
those that do the credible interval is much narrower, as is the case for
\texttt{sx-mathoverflow}. Looking at the shape of the preference
function, there appears to be two distinct shapes of preference
function. The first appears mostly flat (similar to uniform attachment)
for the smallest degrees and then after a threshold linear attachment kicks in, some
with this shape are \texttt{pajek-erdos} and \texttt{sx-mathoverflow}.
The second distinct shape appears to provide some clear linear attachment behaviour
that then slows down after a certain point, and examples of this are
seen in the two infrastructure networks \texttt{opsahl-openflights} and
\texttt{topology}. This slowing down could be viewed as a kind of
diminishing returns on the degree of a node i.e.~as a node gets
larger gaining more connections has less of an effect than it did before
some splice point \(k_0\). 

These empirical insights show the utility of the proposed model for uncovering diverse attachment dynamics in real world networks. In practice, knowing these different behaviors helps us understand how resilient or vulnerable a network is to targeted attacks. When connection growth slows down for large nodes, it prevents a single node from becoming too dominant, making the whole network safer if a main hub fails. On the other hand, knowing when linear growth kicks in helps us spot which networks are becoming overly reliant on a few key hubs and are therefore at greater risk of disruption.

\section{Conclusion and Discussion}\label{sec-conc}

We introduced a class of preference functions that, under the GPA
scheme, generate a network with a flexible yet regularly varying degree
distribution. From the simulation study we showed that the parameters
can be recovered from fitting the model to the degrees alone. We also
applied this method to the degree distributions of real networks,
estimating their model parameters assuming they evolved in the same way.
Not only did this yield good fits for the degree distribution, similar
to that of the IGP mixture, it also came with the added benefit of giving a
posterior estimate for a preference function.

This paper contributes to the understanding of the relationship between
the mechanism underlying a networks growth and the resulting degree
distribution. As well as demonstrating that under certain conditions
information about this mechanism can be garnered from the degrees alone.
In future, something similar could be done by looking at another
statistic for the network (e.g.~the triangle distribution) and using it
in combination with the degrees to gain further insights into the growth
mechanism when the networks full evolution is not available. The
triangle distribution would be a good avenue for future work within the
realm of extremes as it has also been shown by \citet{kang11} that many
real networks seem to have power law triangle distributions.

One limitation of this method is that the lowest degrees needed to be
truncated as they had a very large effect on the fit of the model, as a
result of using theory developed for trees and applying it to general
networks. Future work could apply theory developed for general networks
using a similar method to this, allowing us to compare the results here with
something that is more accurate. This could include fixing the
out-degree of new nodes at a constant greater than one, or allowing the
out-degree of new nodes to vary.

The spliced GPA model relates to bulk-tail approaches in the extremes literature (e.g.\ the EGPD framework of \citet{Papastathopoulos2013}; \citet{Naveau2016}; see also \citet[Ch.~5]{HandbookExtremesCh06}). EGPD-type models allow different behaviour in the body and tail of the distribution itself, chosen for flexibility in fitting shape, with no claim about the underlying process. Our model instead splices the growth rate $b(k)$: nodes attach sub/super-linearly while young, but attachment becomes linear beyond some degree $k_0$. By Theorem~\ref{thm-prp-omega}/Corollary~\ref{thm-omega2}, this asymptotic linearity is exactly the condition for a regularly varying tail, so heaviness is derived from the growth mechanism rather than fitted directly, and $k_0$ estimates where this regime change occurs in the network's growth.

A natural extension would be to let the model parameters depend on
covariates, in the spirit of EGPD regression \citep{deCarvalho2022} or more general extreme value regression \citep{HandbookExtremesCh06}.
This would mean placing covariates on the parameters of the growth
mechanism itself, which would in turn induce an asymptotic form for the tail index $\lambda^*/\beta$ somehow related to
these regression extensions. However, unlike EGPD, where covariates enter the body and tail separately, here $\lambda^*$ depends jointly on the bulk parameters, so we do not expect a one-to-one correspondence between covariate effects on $b(k)$ and covariate effects on the tail index.  What
this correspondence would look like, and whether it can be characterised explicitly, is an interesting open problem that we leave as an avenue for further work.

Also worth noting is the recent work by \citet{banerjee25} that provides
results for more general networks beyond trees utilising the same
underlying branching process. They consider a model that grows in the
exact same way as the model that we have considered, but then they
collapse the nodes of the tree resulting in a more general network much
more like those in reality. However, there is no expression for the
probability mass function of the degrees and therefore no likelihood
that can be used for modelling in the same way that we have. In spite of
this, using Remark 2 from \citet{banerjee25} it can be shown that the
survival function of the limiting degree distribution (when using a
preference function of the form from Theorem~\ref{thm-omega2}) can be
bounded by two regularly varying functions showing that the degree
distribution is still heavy tailed, although not necessarily regularly
varying.

Obtaining an expression for the degree distributions of more general
cases of preferential attachment models are, for the moment, seemingly
inaccessible and provide a real barrier to using more realistic models
in a way similar to what we have in this paper, we leave these open
problems for future analysis.



\section*{References}\label{sec-references}
\addcontentsline{toc}{section}{References}

\renewcommand{\bibsection}{}
\bibliography{refs.bib}

\appendix

\section{Proofs and Derivations}\label{sec-proofs}

\subsection{\texorpdfstring{Proof of
Theorem~\ref{thm-rem-omega}}{Proof of Theorem~}}\label{sec-remproof}

As \(b(k)\) is non-decreasing, when \(b(k)\) is
bounded, by the monotone convergence theorem,
\(\lim_{k\rightarrow\infty}b(k) = c\) for some \(c>0\). Using
Equation~\eqref{eq-surv-origin},
\begin{align*}
    \quad\frac{\bar{F}(k+1)}{\bar{F}(k)}&=\frac{b(k+1)}{\lambda^{*}+b(k+1)}\\
    \quad\lim_{k\rightarrow\infty}\frac{\bar{F}(k+1)}{\bar{F}(k)}&=\frac{c}{\lambda^{*}+c}\\
\quad \lim_{k\rightarrow\infty}\left(\log\frac{\bar{F}(k)}{\bar{F}(k+1)}\right)^{-1}&=\left[\log\left(\frac{\lambda^{*}+c}{c}\right)\right]^{-1} > 0.
\end{align*}
As this means that both terms in Equation~\eqref{eq-omega} have the same finite limit, 
\begin{align*}
\lim_{k\rightarrow\infty}\Omega(F,k) &= \lim_{k\rightarrow\infty}\left(\log\displaystyle\frac{\bar F (k+1)}{\bar F (k+2)}\right)^{-1} - \lim_{k\rightarrow\infty}\left(\log\displaystyle\frac{\bar F (k)}{\bar F (k+1)}\right)^{-1}\\
&=\left[\log\left(\frac{\lambda^{*}+c}{c}\right)\right]^{-1} - \left[\log\left(\frac{\lambda^{*}+c}{c}\right)\right]^{-1} = 0.\qquad\square
\end{align*}
The interpretation is that if $b(k)$ is bounded (above), then the resulting $F$ is not long-tailed, which implies $F$ is light-tailed and recovered to the Gumbel MDA.


\subsection{\texorpdfstring{Proof of
Theorem~\ref{thm-prp-omega}}{Proof of Theorem~}}\label{sec-prpproof}

Taking the form of the GPA degree survival from
Equation~\eqref{eq-surv-origin} : \[
\bar F(k) = \prod_{i=0}^k\frac{b(i)}{\lambda^*+b(i)},
\] and substituting into Equation~\eqref{eq-omega}:
\begin{align*}
\Omega(F,k)&=\left(\log\frac{\prod_{i=0}^{k+1}\frac{b(i)}{\lambda^{*}+b(i)}}{\prod_{i=0}^{k+2}\frac{b(i)}{\lambda^{*}+b(i)}}\right)^{-1}-\left(\log\frac{\prod_{i=0}^{k}\frac{b(i)}{\lambda^{*}+b(i)}}{\prod_{i=0}^{k+1}\frac{b(i)}{\lambda^{*}+b(i)}}\right)^{-1}\\
&=\left(\log\frac{\lambda^{*}+b(k+2)}{b(k+2)}\right)^{-1}-\left(\log\frac{\lambda^{*}+b(k+1)}{b(k+1)}\right)^{-1}\\
&=\left(\log\left[1+\frac{\lambda^{*}}{b(k+2)}\right]\right)^{-1}-\left(\log\left[1+\frac{\lambda^{*}}{b(k+1)}\right]\right)^{-1}\\
&= \left(\log\left[1+\frac{\lambda^{*}}{b(k+2)}\right]\right)^{-1}-\frac{b(k+2)}{\lambda^{*}}+\frac{b(k+2)}{\lambda^{*}}\\&\qquad  -\left(\log\left[1+\frac{\lambda^{*}}{b(k+1)}\right]\right)^{-1}+\frac{b(k+1)}{\lambda^{*}}-\frac{b(k+1)}{\lambda^{*}}\\
&=\left\{ \left(\log\left[1+\frac{\lambda^{*}}{b(k+2)}\right]\right)^{-1}-\frac{b(k+2)}{\lambda^{*}}\right\}\\ &\qquad  - \left\{ \left(\log\left[1+\frac{\lambda^{*}}{b(k+1)}\right]\right)^{-1}-\frac{b(k+1)}{\lambda^{*}}\right\}\\&\qquad  +\frac{b(k+2)}{\lambda^{*}}-\frac{b(k+1)}{\lambda^{*}}.
\end{align*}

Then as \(\lim_{k\rightarrow\infty}b(k)=\infty\) it follows that:
\begin{align*}
\lim_{k\rightarrow\infty}\Omega(F,k) &= \lim_{k\rightarrow\infty}\left\{ \left(\log\left[1+\frac{\lambda^{*}}{b(k+2)}\right]\right)^{-1}-\frac{b(k+2)}{\lambda^{*}}\right\} \\&\qquad - \lim_{k\rightarrow\infty}\left\{ \left(\log\left[1+\frac{\lambda^{*}}{b(k+1)}\right]\right)^{-1}-\frac{b(k+1)}{\lambda^{*}}\right\}\\
&\qquad +\lim_{k\rightarrow\infty}\left(\frac{b(k+2)}{\lambda^{*}}-\frac{b(k+1)}{\lambda^{*}}\right)\\
&=\frac{1}{2}-\frac{1}{2} + \lim_{k\rightarrow\infty}\left(\frac{b(k+2)}{\lambda^{*}}-\frac{b(k+1)}{\lambda^{*}}\right)\\
&=\frac{1}{\lambda^{*}}\lim_{k\rightarrow\infty}\left[b(k+2)-b(k+1)\right].
\end{align*}

The first two lines both have the limit \(1/2\) as
\(\lim_{z\rightarrow{}0}\left[\log(1+z)\right]^{-1}-z^{-1} = 1/2\). This proves the first part of Theorem~\ref{thm-prp-omega}. Now, using Equation~\eqref{eq-surv-origin},
\begin{equation*}\frac{\bar{F}(k+1)}{\bar{F}(k)}=\frac{b(k+1)}{\lambda^{*}+b(k+1)}=\frac{1}{\lambda^{*}/b(k+1) + 1}\rightarrow{}1
\end{equation*}
as \(k\rightarrow\infty\). This means that \(F\) is long-tailed, and
therefore is in an MDA. $\qquad\square$

\subsection{\texorpdfstring{Derivation of
Equation~\eqref{eq-polysurv}}{Derivation of Equation~}}\label{sec-gamma}

For the remaining derivations we rely on a preference function of the form:
\begin{equation}
b(k) = \begin{cases}
g(k),&k\le k_0,\\
g(k_0) + \beta(k-k_0), &k\geq k_0,
\end{cases}\label{eq-g}
\end{equation} for \(\beta>0, k_0\in\mathbb N\), and the identity
\begin{equation}
\prod_{j=0}^{n-1}(x+yj) = x^{n}\frac{\Gamma\left(\frac{x}{y}+n\right)}{\Gamma\left(\frac{x}{y}\right)}. \label{eq-gamma-ratio}
\end{equation} 
Consider the product $\prod_{j=k_0}^{i-1}\frac{b(j)}{\lambda +b(j)}.$
Upon reindexing, applying Equations~\eqref{eq-g} and \eqref{eq-gamma-ratio} and simplifying, we have
\begin{align}
\prod_{j=k_0}^{i-1}\frac{b(j)}{\lambda +b(j)}&=\frac{\prod_{j=k_0}^{i-1}[g(k_0) + \beta(j-k_0)]}{\prod_{j=k_0}^{i-1}[\lambda +g(k_0) + \beta(j-k_0)]}=\nonumber\\
&=\frac{\Gamma\left(\frac{g(k_0)}{\beta}+i-k_0\right)\Gamma\left(\frac{\lambda+g(k_0)}{\beta}\right)}{\Gamma\left(\frac{\lambda+g(k_0)}{\beta}+i-k_0\right)\Gamma\left(\frac{g(k_0)}{\beta}\right)}.\label{eq-gamma-trick}
\end{align}
As the expression for $\bar{F}(k)$ in Equation~\eqref{eq-polysurv} is trivial when $k\leq{}k_0$, we focus on when $k>k_0$ and apply Equation~\eqref{eq-gamma-trick}:
\begin{align*}
\bar{F}(k)&=\prod_{j=0}^{k}\frac{b(j)}{\lambda^{*}+b(j)}=\prod_{j=0}^{k_0-1}\frac{b(j)}{\lambda^{*}+b(j)}\prod_{j=k_0}^{k}\frac{b(j)}{\lambda^{*}+b(j)}\\
&=\prod_{j=0}^{k_0-1}\frac{g(j)}{\lambda^{*}+g(j)}\times\frac{\Gamma\left(\frac{g(k_0)}{\beta}+k-k_0+1\right)\Gamma\left(\frac{\lambda^{*}+g(k_0)}{\beta}\right)}{\Gamma\left(\frac{\lambda^{*}+g(k_0)}{\beta}+k-k_0+1\right)\Gamma\left(\frac{g(k_0)}{\beta}\right)}\\
&=\prod_{j=0}^{k_0-1}\frac{j^{\alpha}+\varepsilon}{\lambda^{*}+j^{\alpha}+\varepsilon}\times\frac{\Gamma\left(\frac{k_0^{\alpha}+\varepsilon}{\beta}+k-k_0+1\right)\Gamma\left(\frac{\lambda^{*}+k_0^{\alpha}+\varepsilon}{\beta}\right)}{\Gamma\left(\frac{\lambda^{*}+k_0^{\alpha}+\varepsilon}{\beta}+k-k_0+1\right)\Gamma\left(\frac{k_0^{\alpha}+\varepsilon}{\beta}\right)}.
\end{align*}

\subsection{\texorpdfstring{Derivation of
Equation~\eqref{eq-rho}}{Derivation of Equation~}}\label{sec-rhoproof}

Using Equations~\eqref{eq-laplace-condition} and \eqref{eq-gamma-trick} and denoting $C_{g,k_0} = \sum_{i=0}^{k_0}\prod_{j=0}^{i-1}\frac{g(j)}{\lambda+g(j)}$,
\begin{align*}
\hat\rho(\lambda) &= \sum_{i=0}^\infty\prod_{j=0}^{i-1}\frac{b(j)}{\lambda+b(j)}\\
&= \sum_{i=0}^{k_0}\prod_{j=0}^{i-1}\frac{b(j)}{\lambda+b(j)} + \sum_{i=k_0+1}^\infty\left(\prod_{j=0}^{k_0-1}\frac{b(j)}{\lambda+b(j)}\prod_{j=k_0}^{i-1}\frac{b(j)}{\lambda +b(j)}\right)\\
&= \sum_{i=0}^{k_0}\prod_{j=0}^{i-1}\frac{g(j)}{\lambda+g(j)} + \sum_{i=k_0+1}^\infty\left(\prod_{j=0}^{k_0-1}\frac{g(j)}{\lambda+g(j)}\prod_{j=k_0}^{i-1}\frac{b(j)}{\lambda +b(j)}\right)\\
&=C_{g,k_0} + \left(\prod_{j=0}^{k_0-1}\frac{g(j)}{\lambda+g(j)}\right)\sum_{i=k_0+1}^\infty\frac{\Gamma\left(\frac{g(k_0)}{\beta}+i-k_0\right)\Gamma\left(\frac{\lambda+g(k_0)}{\beta}\right)}{\Gamma\left(\frac{\lambda+g(k_0)}{\beta}+i-k_0\right)\Gamma\left(\frac{g(k_0)}{\beta}\right)}\\
&= C_{g,k_0} + \frac{\Gamma\left(\frac{\lambda+g(k_0)}{\beta}\right)}{\Gamma\left(\frac{g(k_0)}{\beta}\right)}\left(\prod_{j=0}^{k_0-1}\frac{g(j)}{\lambda+g(j)}\right)\sum_{i=k_0+1}^\infty\frac{\Gamma\left(\frac{g(k_0)}{\beta}+i-k_0\right)}{\Gamma\left(\frac{\lambda+g(k_0)}{\beta}+i-k_0\right)}\\
&=C_{g,k_0} + \frac{\Gamma\left(\frac{\lambda+g(k_0)}{\beta}\right)}{\Gamma\left(\frac{g(k_0)}{\beta}\right)}\left(\prod_{j=0}^{k_0-1}\frac{g(j)}{\lambda+g(j)}\right)\sum_{i=1}^\infty\frac{\Gamma\left(\frac{g(k_0)}{\beta}+i\right)}{\Gamma\left(\frac{\lambda+g(k_0)}{\beta}+i\right)}.
\end{align*}
To simplify the infinite sum, consider:
\begin{align*}
\sum_{i=0}^\infty\frac{\Gamma(i+x)}{\Gamma(i+x+y)} &=\frac{1}{\Gamma(y)}\sum_{i=0}^\infty \text{B}(i+x,y)\\
&=\frac{1}{\Gamma(y)}\sum_{i=0}^\infty\int_0^1t^{i+x-1}(1-t)^{y-1}\text{d}t\\
&=\frac{1}{\Gamma(y)}\int_0^1 t^{x-1}(1-t)^{y-1}\sum_{i=0}^\infty t^i \text{d}t\\
&=\frac{1}{\Gamma(y)}\int_0^1 t^{x-1}(1-t)^{y-1}\frac{1}{1-t}\text{d}t\\
&=\frac{1}{\Gamma(y)}\int_0^1 t^{x-1}(1-t)^{y-2}\text{d}t\\
&=\frac{1}{\Gamma(y)}\text{B}(x,y-1)\\
&= \frac{\Gamma(x)}{(y-1)\Gamma(x+y-1)}.
\end{align*}
This infinite sum does not converge when \(x\le1\) as each term is
\(O(n^{-x})\). We can now use this in \(\hat\rho(\lambda)\):
\begin{align*}
\hat\rho(\lambda) &= C_{g,k_0} + \frac{\Gamma\left(\frac{\lambda+g(k_0)}{\beta}\right)}{\Gamma\left(\frac{g(k_0)}{\beta}\right)}\left(\prod_{j=0}^{k_0-1}\frac{g(j)}{\lambda+g(j)}\right)\\&\qquad \times\left(\frac{\Gamma\left(\frac{g(k_0)}{\beta}\right)}{\left(\frac{\lambda}{\beta}-1\right)\Gamma\left(\frac{g(k_0)+\lambda}{\beta}-1\right)}-\frac{\Gamma\left(\frac{g(k_0)}{\beta}\right)}{\Gamma\left(\frac{g(k_0)+\lambda}{\beta}\right)}\right)\\
&=C_{g,k_0} + \left(\prod_{j=0}^{k_0-1}\frac{g(j)}{\lambda+g(j)}\right)\left(\frac{\Gamma\left(\frac{g(k_0)+\lambda}{\beta}\right)}{\left(\frac{\lambda}{\beta}-1\right)\Gamma\left(\frac{g(k_0)+\lambda}{\beta}-1\right)}-1\right)\\
&=C_{g,k_0} + \left(\prod_{j=0}^{k_0-1}\frac{g(j)}{\lambda+g(j)}\right)\left(\frac{\frac{g(k_0)+\lambda}{\beta}-1}{\frac{\lambda}{\beta}-1}-1\right)\\
&=C_{g,k_0} + \left(\prod_{j=0}^{k_0-1}\frac{g(j)}{\lambda+g(j)}\right)\left(\frac{g(k_0)+\lambda-\beta}{\lambda-\beta}-1\right)\\&=\sum_{i=0}^{k_0}\prod_{j=0}^{i-1}\frac{g(j)}{\lambda+g(j)} + \frac{g(k_0)}{\lambda-\beta}\prod_{j=0}^{k_0-1}\frac{g(j)}{\lambda+g(j)}.
\end{align*}
Substituting $g(j)$ and $g(k_0)$ by $j^{\alpha}+\epsilon$ and $k_0^{\alpha} + \epsilon$, respectively, as according to Equation~\eqref{eq-pref}, arrives at Equation~\eqref{eq-rho}. 

\subsection{\texorpdfstring{Derivation of Equation~\eqref{eq-approx}}{Derivation of Equation~}} \label{sec-approx}

Denote the conditional limiting survival of the spliced GPA model by \(\bar F (k|k\ge k_0)\), then:
\begin{align*}
\bar F(k|k>k_0) &= \frac{\bar F(k)}{\bar F(k_0)}\\
&= \frac{\lambda^* + k_0^\alpha + \varepsilon}{k_0^\alpha + \varepsilon}\times \frac{\Gamma\left(\frac{\lambda^*+k_0^\alpha + \varepsilon}{\beta}\right)}{\Gamma\left(\frac{k_0^\alpha + \varepsilon}{\beta}\right)} \frac{\Gamma\left(k-k_0 + 1 +\frac{k_0^\alpha + \varepsilon}{\beta}\right)}{\Gamma\left(k-k_0 + 1 +\frac{\lambda^* +k_0^\alpha + \varepsilon}{\beta}\right)}\\
&=\frac{\Gamma\left(1 + \frac{\lambda^*+k_0^\alpha + \varepsilon}{\beta}\right)}{\Gamma\left(1+\frac{k_0^\alpha + \varepsilon}{\beta}\right)} \frac{\Gamma\left(k-k_0 + 1 +\frac{k_0^\alpha + \varepsilon}{\beta}\right)}{\Gamma\left(k-k_0 + 1 +\frac{\lambda^* +k_0^\alpha + \varepsilon}{\beta}\right)}\\
&\approx \left(1+\frac{k_0^\alpha + \varepsilon}{\beta}\right)^{\lambda^*/\beta} \left(k-k_0+1+\frac{k_0^\alpha + \varepsilon}{\beta}\right)^{-\lambda^*/\beta}\\
&=\left(\frac{\beta(k-k_0)}{k_0^{\alpha}+\varepsilon+\beta} + 1\right)^{-\lambda^{*}/\beta}.
\end{align*}
The approximation in the second last line follows from $\Gamma(z+a)/\Gamma(z+b)\approx{}z^{a-b}$ for large $z$ \citep{gammaratio}.

\subsection{\texorpdfstring{Derivation of Equation~\eqref{eq-likelihood}}{Derivation of Equation~}} \label{sec-likelihood}

To derive the likelihood of the spliced GPA model for data truncated by $l\geq0$, we first derive the \textit{conditional} pmf, denoted by $f_l(k)$, for a general preference function. For $k\geq{}l$,
\begin{align*}
  f_l(k)=\frac{f(k)}{\bar{F}(l-1)}&=\frac{\lambda^{*}}{\lambda^{*}+b(k)}\prod_{i=0}^{k-1}\frac{b(i)}{\lambda^{*}+b(i)}\left/\prod_{i=0}^{l-1}\frac{b(i)}{\lambda^{*}+b(i)}\right.\\
  &=\frac{\lambda^{*}}{\lambda^{*}+b(k)}\prod_{i=l}^{k-1}\frac{b(i)}{\lambda^{*}+b(i)}.\nonumber
\end{align*}
For notational convenience, we denote the likelihood simply by $L$. Using Equation~\eqref{eq-g} and the relevant degree counts $n_l,n_{l+1},\ldots,n_M$,
\begin{align*}
L&=\prod_{i=l}^{M}f_l(i)^{n_i}
=\prod_{i=l}^{M}\left[\frac{\lambda^{*}}{\lambda^{*}+b(i)}\prod_{j=l}^{i-1}\frac{b(j)}{\lambda^{*}+b(j)}\right]^{n_i}\\
&=\prod_{i=l}^{k_0}\left[\frac{\lambda^{*}}{\lambda^{*}+b(i)}\prod_{j=l}^{i-1}\frac{b(j)}{\lambda^{*}+b(j)}\right]^{n_i}\\
&\quad\times\prod_{i=k_0+1}^{M}\left[\frac{\lambda^{*}}{\lambda^{*}+b(i)}\prod_{j=l}^{k_0-1}\frac{b(j)}{\lambda^{*}+b(j)}\prod_{j=k_0}^{i-1}\frac{b(j)}{\lambda^{*}+b(j)}\right]^{n_i}\\
&=\prod_{i=l}^{k_0}\left[\frac{\lambda^{*}}{\lambda^{*}+b(i)}\prod_{j=l}^{i-1}\frac{b(j)}{\lambda^{*}+b(j)}\right]^{n_i}\times\prod_{i=k_0+1}^{M}\left[\prod_{j=l}^{k_0-1}\frac{b(j)}{\lambda^{*}+b(j)}\right]^{n_i}\\
&\quad\times\prod_{i=k_0+1}^{M}\left[\frac{\lambda^{*}}
{\lambda^{*}+b(j)}\prod_{j=k_0}^{i-1}\frac{b(j)}{\lambda^{*}+b(j)}\right]^{n_i}
\end{align*}
Using Equation~\eqref{eq-gamma-trick} product within the square brackets in the second line,
\begin{align*}
L&=\prod_{i=l}^{k_0}\left[\frac{\lambda^{*}}{\lambda^{*}+g(i)}\prod_{j=l}^{i-1}\frac{g(j)}{\lambda^{*}+g(j)}\right]^{n_i}\times\left(\prod_{j=l}^{k_0-1}\frac{g(j)}{\lambda^{*}+g(j)}\right)^{\sum_{i=k_0+1}^{M}n_i}\\
&\quad\times\prod_{i=k_0+1}^{M}\left[\frac{\lambda^{*}}{\lambda^{*}+g(k_0)+\beta(i-k_0)}\times\frac{\Gamma\left(\frac{g(k_0)}{\beta}+i-k_0\right)\Gamma\left(\frac{\lambda^{*}+g(k_0)}{\beta}\right)}{\Gamma\left(\frac{\lambda^{*}+g(k_0)}{\beta}+i-k_0\right)\Gamma\left(\frac{g(k_0)}{\beta}\right)}\right]^{n_i}\\
&=\prod_{i=l}^{k_0}\left[\frac{\lambda^{*}}{\lambda^{*}+i^{\alpha}+\varepsilon}\prod_{j=l}^{i-1}\frac{j^{\alpha}+\varepsilon}{\lambda^{*}+j^{\alpha}+\varepsilon}\right]^{n_i}\times\left(\prod_{j=l}^{k_0-1}\frac{j^{\alpha}+\varepsilon}{\lambda^{*}+j^{\alpha}+\varepsilon}\right)^{\sum_{i=k_0+1}^{M}n_i}\\
&\quad\times\prod_{i=k_0+1}^{M}\left[\frac{\frac{\lambda^{*}}{\beta}}{\frac{\lambda^{*}+k_0^{\alpha}+\varepsilon}{\beta}+i-k_0}\times\frac{\Gamma\left(\frac{k_0^{\alpha}+\varepsilon}{\beta}+i-k_0\right)\Gamma\left(\frac{\lambda^{*}+k_0^{\alpha}+\varepsilon}{\beta}\right)}{\Gamma\left(\frac{\lambda^{*}+k_0^{\alpha}+\varepsilon}{\beta}+i-k_0\right)\Gamma\left(\frac{k_0^{\alpha}+\varepsilon}{\beta}\right)}\right]^{n_i}\\
&=\prod_{i=l}^{k_0}\left[\frac{\lambda^{*}}{\lambda^{*}+i^{\alpha}+\varepsilon}\prod_{j=l}^{i-1}\frac{j^{\alpha}+\varepsilon}{\lambda^{*}+j^{\alpha}+\varepsilon}\right]^{n_i}\times\left(\prod_{j=l}^{k_0-1}\frac{j^{\alpha}+\varepsilon}{\lambda^{*}+j^{\alpha}+\varepsilon}\right)^{\sum_{i=k_0+1}^{M}n_i}\\
&\quad\times\prod_{i=k_0+1}^{M}\left[\frac{\lambda^{*}\,\Gamma\left(\frac{k_0^{\alpha}+\varepsilon}{\beta}+i-k_0\right)\Gamma\left(\frac{\lambda^{*}+k_0^{\alpha}+\varepsilon}{\beta}\right)}{\beta\,\Gamma\left(\frac{\lambda^{*}+k_0^{\alpha}+\varepsilon}{\beta}+i-k_0+1\right)\Gamma\left(\frac{k_0^{\alpha}+\varepsilon}{\beta}\right)}\right]^{n_i}.
\end{align*}

\end{document}